\documentclass[iop]{emulateapj-rtx4} % read by: ACROREAD
\shortauthors{Sekanina}
\shorttitle{Comet C/1995 O1:\ Nucleus and Its Fragmentation}
\slugcomment{Version \today }

\newcommand{\gapeq}{$\;$\raisebox{0.3ex}{$>$}\hspace{-0.28cm}\raisebox{-0.75ex}{$\sim$}$\;$}

\begin{document}
\title{ORBITAL EVOLUTION, ACTIVITY, AND MASS LOSS OF COMET C/1995 O1
(HALE-BOPP).\\II.\ NUCLEUS AND COMPANIONS AS COMPACT CLUSTERS OF MASSIVE FRAGMENTS}
\author{Zdenek Sekanina}%$^1$ \& Rainer Kracht$^2$}
\affil{Jet Propulsion Laboratory, California Institute of Technology,
  4800 Oak Grove Drive, Pasadena, CA 91109, U.S.A.}%\\
% $^2$Ostlandring 53, D-25335 Elmshorn, Schleswig-Holstein, Germany}
\email{Zdenek.Sekanina@jpl.nasa.gov}
% {\hspace*{2.59cm}}R.Kracht@t-online.de{\vspace{-0.1cm}}}

\begin{abstract}
The prime objective is to settle a contradiction between a high nongravitational
acceleration~\mbox{affecting} the orbital motion of comet C/1995 O1 and its enormous
nucleus by modeling it as a {\it compact~cluster~of boulder-sized fragments\/} held
together by its own gravity.  The nongravitational effect is interpreted~as~a
perturbation of the cluster's principal, most massive fragment.  This and other
constraints suggest~that the principal fragment was probably 8--9~km across and the
entire cluster $\sim$150~times~less~\mbox{massive}~than a~\mbox{single-body}~\mbox{nucleus}~of~an~equal~\mbox{cross-sectional}~area~\mbox{derived}~from~the~\mbox{Herschel}~\mbox{far-infrared}~\mbox{photometry}
of the inactive comet detected near 30~AU from the Sun. The cross-sectional area
required~the~\mbox{smallest} fragments to be a few tens of meters across under a
steady-state size distribution.  The cluster~was~at most $\sim$200 km in diameter,
subject to frequent collisions and significant perturbations by~the~Sun~near
perihelion, and apparently a product of tidal fragmentation of the original
nucleus, more than 20~km across, at the time of close encounter with Jupiter
4 millennia ago, if the comet's tensile~strength~was then as low as several Pa.
Published for the first time are the results of a search for~\mbox{companion}~\mbox{nuclei} in
three post-perihelion images taken with the HST's STIS instrument~in~1997--1998.
At~least~29~such objects $<$1200~km (projected) from the primary were detected,
with their signals from 2.3\% to 25\%~of the primary's and the signal-to-noise
ratios between 5:1 and 29:1, apparently also~cluster-like~in~nature.
\end{abstract}


\section{Introduction}
In Part I (Sekanina \& Kracht 2017; hereafter Paper~1) the orbit of comet C/1995
O1 was determined from astrometric observations covering an arc of 17.6~yr; the
conditions were examined that existed at the time of a predicted close encounter
with Jupiter on $-$2251 November 7; a correlation was established between a
nongravitational acceleration in the comet's orbital motion and the mass loss
rates of water and other ices sublimated~from the nucleus; and the production
and dynamics of dust ejecta were addressed.  Emphasized was the role of non-water
compounds --- low-volatility organic molecules in particular --- in the process
of outgassing and their contribution to the total outgassed mass, and also noted
was a high mass loading of the gas flow from the nucleus by the released dust.
Finally, the issue was brought up of a major disparity between the comet's mass
determined dynamically vs photometrically, which is critical for a makeup of the
nucleus, the objective of this study.

\section{Nongravitational Effect on the Nucleus and Conservation-of-Momentum Law}
% \section{Nongravitational Effect and Reported Size of the Nucleus}
%
It was briefly remarked in Paper 1 that the~\mbox{magnitude} of the
nongravitational acceleration in the motion of C/1995 O1 was equivalent, after
integrating over the orbital period, to a momentum change per unit mass of
\mbox{2.46$\,\pm\,$0.14~m~s$^{-1}$}.  This result was derived by optimizing a
modified version of Marsden et al.'s (1973) nongravitational law, whose scaling
distance of 15.36~AU was determined, by fitting 1950 astrometric observations from
1993--2010, as part of a preferred orbital solution.~Fully 80\% of the effect
was contributed by the component of the nongravitational acceleration that was
directed away from the Sun and long recognized as the primary trigger of the
outgassing-driven recoil motion of the nucleus, invoked by a
conservation-of-momentum law (e.g., Bessel 1836; Whipple 1950; Marsden 1968, 1969).

For a given position in a comet's orbit, at time $t$,~the conservation-of-momentum
law is commonly written as a one-dimensional condition.  If only a single species
sublimates, the condition reads
\begin{equation}
{\cal M} \, \zeta_{\rm ng}(t) + \alpha_{\rm gas}(t) \, \dot{m}_{\rm sub}(t) \,
 v_{\rm sub}(t) = 0,
\end{equation}
where ${\cal M}$ is the nucleus' mass and, at time $t$,~\mbox{$\zeta_{\rm ng}(t)
> 0$} the nongravitational acceleration, \mbox{$\dot{m}_{\rm sub}(t) < 0$} the
mass-loss rate by outgassing (i.e., the mass production~rate~of gas), $v_{\rm
sub}$ the velocity with which the gas sublimates, and \mbox{$\alpha_{\rm gas}(t)
< 1$} a recoil parameter that accounts for a vectorial distribution of the
outgassed mass, whereby only a fraction of the momentum generated by the
sublimating ice is, in general, transformed into the detected nongravitational
acceleration.  The recoil parameter depends on the surface distribution of the ice,
the nucleus' shape and rotation, the gas-flow collimation, etc.  When a number of
different species, $n_{\rm gas}$, sublimate at the same time, the second term in
{\vspace{-0.05cm}} Equation~(1) is to be replaced with a sum \mbox{$\Sigma_{i=1}
^{n_{\rm gas}} \,\alpha_{{\rm gas},i}\,\dot{m}_{{\rm sub},i}\,v_{{\rm sub},i}$}.
Because $v_{{\rm sub},i}$ is a function of the mass loading of the gas flow by
dust, {\vspace{-0.03cm}} it could depend not only on $\dot{m}_{{\rm sub},i}$, but
on \mbox{$\dot{m}_{{\rm sub},j}$ ($j \!=\!1, \ldots, n_{\rm gas}, j \!\neq\! i)$}
as well.  The mass of the nucleus is assumed not to vary with time in Equation~(1),
because \mbox{$\dot{m}_{\rm sub} \, dt \!\ll\! {\cal M}$}.

The conservation-of-momentum condition may also be integrated over the entire
orbit, specifically,
\begin{equation}
{\cal M} \!\! \int_{t_\pi-\frac{1}{2}P}^{t_\pi+\frac{1}{2}P} \!\!\!\!
 \zeta_{\rm ng}(t) \, dt + \langle \alpha_{\rm gas} v_{\rm sub} \rangle \!
 \sum_{i=1}^{n_{\rm gas}} \! \int_{t_\pi-\frac{1}{2}P}^{t_\pi+\frac{1}{2}P}
 \!\!\!\! \dot{m}_{{\rm sub},i}(t) \, dt = 0,
\end{equation}
where $t_\pi$ is the perihelion time, $P$ the orbital period, and $\langle
\alpha_{\rm gas} v_{\rm sub} \rangle$ is an orbit-averaged value of the product
of the recoil parameter $\alpha_{\rm gas}$ and the sublimation velocity $v_{\rm
sub}$.  While the integrated expressions were determined in Paper~1, the product
$\langle \alpha_{\rm gas} v_{\rm sub} \rangle$ is subject to some uncertainty,
but can be constrained as follows.

For any particular ice, the initial velocity of sublimation is subsonic in the
presence of dust (Probstein 1969).  Hence, if $v_{\rm son}$ is the speed of
sound in the gas flow, then
\begin{equation}
v_{\rm sub}(t) = \beta_{\rm gas}(t) \, v_{\rm son}(t) = \beta_{\rm gas}(t)
 \sqrt{\frac{\Re \, \gamma_{\rm gas} \, T_{\rm sub}(t)}{\mu_{\rm gas}}},
\end{equation}
where \mbox{$\beta_{\rm gas} \!<\! 1$} depends on the mass{\vspace{-0.03cm}}
loading of the gas flow by dust, $\Re$ is the gas constant
(8.31\,J\,K$^{-1}$mol$^{-1}$)~and $T_{\rm sub}$, $\gamma_{\rm gas}$, and
$\mu_{\rm gas}$ are, respectively, the~gas~\mbox{temperature} at sublimation,
the heat-capacity ratio, and the molar mass of the species.  Because of a rapid
drop in the mass-loss rate with heliocentric distance, the greatest weight in
the expression{\vspace{-0.05cm}} for $\langle \alpha_{\rm gas} v_{\rm sub}
\rangle$ have the values near peri\-helion.~For water, \mbox{$\gamma_{_{{\rm
H}_2{\rm O}}} \!=\! 1.33$}, \mbox{$\mu_{_{{\rm H}_2{\rm O}}} \!=\! 18.02$ g
mol$^{-1}$},~and{\vspace{-0.06cm}} for C/1995~O1 at perihelion --- employing an
isothermal model of Paper~1 --- a sonic flow (Mach number of \mbox{${\sf M} =
1$}) has a sublimation temperature of \mbox{$T_{{\rm sub},{\scriptscriptstyle
{\rm H}_2{\rm O}}} = 176$\,K} {\vspace{-0.04cm}}and, accordingly, \mbox{$v_{{\rm
son},{\scriptscriptstyle {\rm H}_2{\rm O}}} \simeq 330$ m s$^{-1}$}.

For carbon monoxide the problem is more complex because of a possible effect of
superheating (Fulle et al.\ 1998).  While from the isothermal model one finds that
at perihelion \mbox{$v_{{\rm son},{\scriptscriptstyle {\rm CO}}} \simeq 130$\,K},
Biver et al.'s (2002) monitoring program of the CO kinetic temperature in
C/1995~O1 between perihelion and some \mbox{6--8}~AU both before and after
perihelion suggests \mbox{$T_{{\rm kin},{\scriptscriptstyle {\rm CO}}} \!=\! 113
\pm 6$\,K} {\vspace{-0.04cm}}at perihelion.  Since \mbox{$T_{\rm kin} \!>\!
T_{\rm sub}$}, this implies that \mbox{$v_{{\rm son},{\scriptscriptstyle {\rm
CO}}} \!< 220$ m s$^{-1}\!$}.

For other ices the isothermal~model~offers~for~the~speed of sound numbers that
are~\mbox{intermediate}~\mbox{between}~\mbox{carbon} monoxide and water.  In
particular,~for~\mbox{complex}~\mbox{organic} molecules appropriate
estimates\footnote{Statistically, numerous hydrocarbons appear to satisfy a
relation between the molar mass (in g~mol$^{-1}$) and the heat-capacity ratio
that is approximately expressed as \mbox{$\gamma_{\rm gas} \!=\!  1.56 \!-\!
0.25 \log \mu_{\rm gas}$} for \mbox{$30 \!<\! \mu_{\rm gas} \!<\!  120$ g
mol$^{-1}$}.}\,are \mbox{$T_{\rm sub} \simeq 270$--280\,K}, \mbox{$\mu_{\rm gas}
\!\simeq\! 70\:$g$\: $mol$^{-1}\!$}, and \mbox{$\gamma_{\rm gas} \!\simeq\!
1.1$}, thus \mbox{$v_{\rm son} \!\simeq\! 190$\,m\,s$^{-1}\!$}.

For the speed of sound I now accept a representative value of \mbox{$\langle
v_{\rm son} \rangle = 270$ m s$^{-1}$}, a compromise between water and the
other parent molecules.  With two basic quantities that enter Equation~(2)
--- the integrated nongravitational effect,
\begin{equation}
\Delta V_{\rm ng} = \!\! \int_{t_\pi-\frac{1}{2}P}^{t_\pi+\frac{1}{2}P} \!\!\!
 \zeta_{\rm ng}(t) \, dt = 2.46 \pm 0.14\;{\rm m}\;{\rm s}^{-1}, %1.97, 0.16
\end{equation}
and the integrated mass loss of water ice,
\begin{equation}
\Delta {\cal M}_{\scriptscriptstyle {\rm H}_2{\rm O}} = \!\!
 \int_{t_\pi-\frac{1}{2}P}^{t_\pi+\frac{1}{2}P} \!\!\! \dot{m}_{{\rm sub},
 {\scriptscriptstyle {\rm H}_2{\rm O}}}(t) \, dt =
 -3.4\:\!\!\times\:\!\!\!10^{15}\,{\rm g}
\end{equation}
--- already determined in Paper 1, I next turn to a total orbit-integrated mass
loss by outgassing.  Examination of the contributions from a large set of non-water
species resulted in Paper~1 in a total {\it documented\/}~mass~loss~of~175\% of
the loss of water and a {\it predicted\/} range of total~losses well over 200\%,
a conclusion based primarily on a recognition of an apparently highly incomplete
inventory of complex organic molecules.  \mbox{Crovisier}~et~a.~(2004)~similarly
argued that there were still many molecular species to be discovered in
comets.~However, rather~than the orbit-integrated mass-loss data they used
near-perihelion abundances, in which case the degree of incompleteness --- while
still detectable --- appears to be less prominent.  Based on the results of
Paper~1, I adopt for the total orbit-integrated mass loss by outgassing a
representative value of 250\% of the mass loss by water ice:
\begin{eqnarray}
\Delta {\cal M}_{\rm gas} & = & \sum_{i=1}^{n_{\rm gas}}
 \int_{t_\pi-\frac{1}{2}P}^{t_\pi+\frac{1}{2}P} \!\!\!
 \dot{m}_{{\rm sub},i}(t) \, dt 
 \nonumber \\[-0.36cm]
 & & \\[-0.06cm]
 & = & 2.5 \Delta {\cal M}_{\scriptscriptstyle {\rm H}_2{\rm O}}
 = -8.5\:\!\!\times\:\!\!\!10^{15}\,{\rm g}. \nonumber
\end{eqnarray}
Equation (2), in which --- following (3) --- $\langle \alpha_{\rm gas} \,
v_{\rm sub} \rangle$ is replaced with $\langle \alpha_{\rm gas} \,
\beta_{\rm gas} \, v_{\rm son} \rangle$, can be used to estimate an
upper limit on the mass of the nucleus by substituting $\langle v_{\rm son}
\rangle$ for this expression:
\begin{equation}
{\cal M} = \langle \alpha_{\rm gas} \, \beta_{\rm gas} \, v_{\rm son} \rangle
 \frac{|\Delta {\cal M}_{\rm gas}|}{\Delta V_{\rm ng}} < \langle v_{\rm son}
 \rangle \frac{|\Delta {\cal M}_{\rm gas}|}{\Delta V_{\rm ng}} .
\end{equation}
The inequality follows from $\alpha_{\rm gas}$ and $\beta_{\rm gas}$ being
always smaller than unity.  This relation indicates that since the sublimation
velocity amounted to less than the speed of sound~and the gas flow was
imperfectly collimated, the momentum of the outgassed mass per orbit should
have been less than \mbox{2.3$\,\times$10$^{20}$\,g cm s$^{-1}$} and the
nucleus less than \mbox{1$\,\times$10$^{18}$\,g} in mass.  Assuming a bulk
density of 0.4~g~cm$^{-3}$, the diameter should under no~\mbox{circumstances}
have exceeded 17~km.  In reality, the heavy mass loading of the gas flow by dust,
which --- based on the results of Paper~1 {\it and\/} including the contributions
from as yet undetected molecules, as implied by Equation~(6) --- is likely to
have exceeded~4, suggests an initial Mach number of \mbox{${\sf M} < 0.3$} for
the gas (Probstein~1969), while the absence of perfect gas-flow collimation
may have reduced effects of the momentum by another factor of 1.5 or more.
And even though there may have existed phenomena that worked in the opposite
direction (such as recondensation; fallback on the surface by boulders; etc.),
the momentum imparted to the nucleus should still have been substantially lower
than implied by $\langle v_{\rm son}\rangle$.  Allowing~the~product of
\mbox{$\langle \alpha_{\rm gas} \beta_{\rm gas} \rangle$} to vary from 0.1
to a stretched value of 0.35, the diameter of a model spherical nucleus
consistent with Equation~(7) should be in a range from $\sim$8~km to $\sim$12~km.

Szab\'o et al.\ (2011, 2012) analyzed several images of C/1995 O1 along the
post-perihelion leg of the orbit up to a heliocentric distance of 32~AU.  The
authors concluded that the comet's activity ceased between late 2007 and early
2009; this estimate can be refined with use of the results by Kramer et al.\
(2014), who still detected minor activity in 2008 August-September.  The
inactive nucleus was detected at optical wavelengths of 0.55--0.9~$\mu$m on
a few occasions, including with the Hubble Space Telescope (HST) on 2009
September 8 and with the Very Large Telescope (VLT) on 2011 October 5--25.
It was also observed with the Herschel Space Observatory at 70~$\mu$m on
2010 June 10.  These observations allowed Szab\'o et al.\ (2012) to determine
separately a cross-sectional area of the nucleus, yielding a mean diameter of
\mbox{74$\,\pm\,$6}~km with an axial ratio of at least \mbox{1.72$\,\pm\,$0.07};
and a post-perihelion geometric albedo of \mbox{0.081$\,\pm\,$0.009}, in
contrast to a preperihelion albedo of only \mbox{0.03$\,\pm\,$0.01}.

Szab\'o et al.'s (2012) determination of the nucleus' dimensions is in remarkably
good agreement with an earlier result by Sekanina (1999a), who derived an average
diameter of \mbox{71$\,\pm\,$4 km} (cf.\ Table~5) from six HST images exposed
between October 1995 and October 1996 (i.e., long before perihelion).  Other
researchers arrived at less consistent dimensions, commenting on large
discrepancies among the determinations by diverse methods.  Employing the same
HST images but a different approach, Weaver \& Lamy (1999) did estimate the
most probable diameter at 70~km, yet not entirely ruling out \mbox{30--40 km};
they also reviewed the results by another group of an occultation of a star by
the comet and arrived at an admittedly model dependent diameter of less than
52~km, while from three independent microwave observations the diameter came
out to be near 40~km. Thermal-infrared observations aboard the Infrared Space
Observatory resulted in a diameter of \mbox{70--112}~km depending on the applied
physical model (Jorda et al.\ 2000).  Re-reviewing the constraints from these
data and near-perihelion radiometric data, Fern\'andez (2002) estimated the
diameter at \mbox{60$\,\pm\,$20 km}.  In their summary table, Lamy et al.\
(2004) provide two numbers for the nucleus' diameter, 74~km and 60~km, with
no errors provided.

In summary, the {\it dimensions of an inactive nucleus\/} of C/1995~O1 derived
for a single-body model from a~far-infrared\mbox{\hspace{0.07cm}}observation\mbox{\hspace{0.07cm}}at\mbox{\hspace{0.07cm}}a\mbox{\hspace{0.07cm}}record\,large\,heliocentric\,distance do {\it under no circumstances accommodate the magnitude of
the outgassing-driven nongravitational acceleration\/} in the comet's motion.
In the conservation-of-momentum equation, this disparity exceeds two orders of
magnitude in terms of the mass of the nucleus, so that one confronts a {\it major
contradiction\/}.  Interestingly, this problem was independently mentioned by Sosa
\& Fern\'andez (2011), yet it has never been solved nor its possible implications
seriously addressed in the literature.

\section{Nucleus of C/1995 O1 as a Compact Cluster of Fragments of the Original
 Body}
I argue that the only feasible solution to this contention requires one to
postulate that the nucleus of C/1995~O1 at its recent return to perihelion
was being made up of~a {\it compact cluster of massive fragments\/} into which the
original nucleus broke up by the action of tidal forces exerted by Jupiter
in the course of the comet's close encounter with the planet in the 3rd
millennium BCE (Paper~1).

I further postulate that the detected nongravitational acceleration refers to
the principal, most massive fragment, \mbox{8--12}~km in diameter
and orbiting the Sun close to the cluster's center of mass, while the
nongravitational accelerations~on other outgassing fragments trigger minor
perturbations of their motions relative to the principal fragment and
remain undetected.

For this model to be physically meaningful, the cluster must be bound enough
by gravity to survive as a densely-packed assemblage over more than 4000~yr
against both the Sun's perturbations and collisional self-destruction.  It is
also necessary to account for the observed nucleus' brightness in terms of a
cross-sectional area of the (optically thin) cloud of fragments, to establish
their size range and distribution to make the model self-consistent, and to
describe the fragment properties in the context of both observations and
constraints on the model --- including its gravitational stability and
collisional history.

\section{Stability of Gravitationally Bound Orbits:\ The Analogs}
Globular clusters come to mind as a convenient cosmic analog for investigating
the conditions of gravity-driven stability of a compact cometary assemblage.
Kennedy's (2014) recent paper extensively deals with this issue and predicts
the radius of stability, $r_{\rm st}$, employing Mardling's (2008) analysis.
The outcome is a simple expression
\begin{equation}
r_{\rm st} = r_{\rm gal} f_0 \! \left( \!\frac{{\cal M}_{\rm gc}}{{\cal M}_{\rm
 gal}} \! \right)^{\!\!\frac{1}{3}} \!\!,
\end{equation}
where ${\cal M}_{\rm gal}$, ${\cal M}_{\rm gc}$, $r_{\rm gal}$, and $f_0$ are,
respectively, the mass of the galaxy, the mass of the globular cluster, the
distance of its closest approach to the center of the galaxy, and a parameter
that is a function of the globular cluster's  orbital eccentricity.  In a
limiting parabolic scenario, the orbits of all stars in the cluster are stable
when \mbox{$f_0 < 0.18$}.  For comparison, the radius of Hill's sphere~of
C/1995~O1 at time $t$ is (e.g., Chebotarev 1964)
\begin{equation}
r_{\rm Hill}(t) = r_{\rm st}(t,h_0) = r(t) \, h_0 \! \left( \! \frac{{\cal
 M}_{\rm C}}{{\cal M}_{\rm Sun}} \! \right)^{\!\!\frac{1}{3}} \!\!,
\end{equation}
where \mbox{$h_0 = 3^{-\frac{1}{3}} = 0.69$} and ${\cal M}_{\rm Sun}$, ${\cal
M}_{\rm C}(t)$, and $r(t)$ are, respectively, the mass of the Sun, and the
comet's mass and heliocentric distance at time $t$.  Kennedy's (2014) results
show that the stability-zone's radius is about four or more times smaller than
the radius of Hill's sphere.

Two papers by Hamilton \& Burns (1991, 1992) on orbital stability zones about
asteroids are highly relevant.  The authors fortunately extended the range of
investigated orbital eccentricities to 0.9, thereby covering effec\-tively comets
as well.  Establishing the stability of gravitationally bound orbits of particles
around asteroids as a function of a deviation from orbit circularity, they
also considered effects of the Coriolis force, orbital inclination, and solar
radiation pressure.   They further showed that escape to interplanetary space was
not the only loss mechanism for particles in the initially bound trajectories;
impact~on the asteroid's surface was another means of removal.  Their effort
focused on the determination of a ratio equivalent to $D_{\rm st}/D_{\rm C}$,
where $D_{\rm st}$ is the stability-zone's diameter (equaling $2r_{\rm st}$)
and $D_{\rm C}$ is the comet's or asteroid's diameter.  For elongated orbits
they independently concluded that the heliocentric distance in Equation~(9)
was to be taken at perihelion, \mbox{$r(t_\pi) \!=\! q$}, so that
\begin{equation}
\frac{D_{\rm st}}{D_{\rm C}} = 191.76 h_0 q \rho^{\frac{1}{3}},
\end{equation}
where $\rho$ is the bulk density of the comet's nucleus or the asteroid in
g~cm$^{-3}$ and $q$ is in AU. 

Hamilton \& Burns (1991) demonstrated that the loss~of orbital stability was
considerably higher among particles in prograde than in retrograde trajectories.
In fact, extrapolation to the parent body's parabolic orbit would leave practically
no gravitationally bound particles after 20~yr, the period of time over which
their integrations extended.  For retrograde orbits, impacts on the asteroid's
surface became extremely rare and extrapolation of the stability-zone's radius to
parabolic motion was uncertain, in part because of insufficient integration
times, but some particles were likely to have survived in bound orbits.
Estimating crudely from Hamilton \& Burns' plot that \mbox{$D_{\rm st}/D_{\rm C}
> 10$}, one obtains \mbox{$h_0 > 0.08$} from Equation~(10) with \mbox{$\rho =
0.4$ g cm$^{-3}$}.  Combining this lower limit with the upper bound implied by
Kennedy's (2014) computations, I adopt in the following \mbox{$h_0 = 0.1$} to
derive $D_{\rm st}$ for C/1995~O1.

Hamilton \& Burns (1992) found that radiation pressure eliminated from the
stable orbits all particles smaller than $\sim$1~mm for an asteroid $\sim$200~km
in diameter and all particles smaller than $\sim$1~cm for an asteroid $\sim$20~km
in diameter.  There was little difference between radiation-pressure effects on
prograde and retrograde orbits.  Based on these results, it can be expected that
no fragments smaller than a few centimeters would survive in stable orbits in
the presumed nucleus' cluster of C/1995~O1.

The results by Hamilton \& Burns (and others referred to in their papers) were
a product of solving a \mbox{three-body} problem:\ the Sun, an asteroid, and
a single orbiting particle.  By contrast, a cluster of fragments involves of
course an $n$-body problem, as described below.

\section{Formulation of a Compact Cluster Model}
To describe the compact cluster of fragments, it is desirable to first constrain
its properties by requiring that they satisfy relevant observations.  If a
product of~a collisional process that began with an initial tidal breakup in
close proximity of Jupiter and was characterized by very low relative
velocities (Section 8), one expects the differential size distribution function
of fragments, i.e., their number, $d{\cal N}_{\rm frg}$, with diameters from $D$
to \mbox{$D \!+\! dD$}, to eventually approach steady state.  At a constant
bulk density, this scenario implies (Dohnanyi 1969; Williams \& Wetherill 1994)
\begin{equation}
d{\cal N}_{\rm frg}(D) = k_{\rm frg} D^{-\frac{7}{2}} dD,
\end{equation}
where $k_{\rm frg}$ is a normalization constant.  The cumulative distribution,
i.e., the number of fragments whose diameters are greater than, or equal to, $D$,
is then
\begin{equation}
{\cal N}_{\rm frg}(D) = \!\!\int_{D}^{\infty}\!\!\!k_{\rm frg} D^{-\frac{7}{2}}\,dD
 = \frac{2k_{\rm frg}}{5} D^{-\frac{5}{2}}.
\end{equation}
For the principal fragment, whose diameter $D_0$ was constrained by the
orbital-momentum condition (Sections~2 and 3) to a range of \mbox{8--12}~km,
{\vspace{-0.09cm}}this expression requires that \mbox{${\cal N}_{\rm frg}(D_0)
\!=\! 1$}, so that \mbox{$k_{\rm frg} \!=\! \frac{5}{2} D_0^{5/2}$} and
Equation~(12) simplifies to
\begin{equation}
{\cal N}_{\rm frg}(D) = \!\left(\!\frac{D_0}{D} \!\right)^{\!\!\frac{5}{2}} \!\!.
\end{equation}
If, reckoned in the order of decreasing size, an $i$-th fragment (${\cal N}_{\rm
frg} \!=\! i$) has a diameter $D_{i-1}$, the diameter $D_i$ of the next smaller,
\mbox{$(i\!+\!1)$}st fragment equals
\begin{equation}
D_i = D_{i-1}\!\left[ 1 \!+\! \frac{1}{{\cal N}_{\rm frg}(D_{i-1})}
 \right]^{\!-\frac{2}{5}} \!\! = D_{i-1} \! \left( \frac{i}{i\!+\!1} \!
 \right)^{\!\!\frac{2}{5}}\!\!.
\end{equation}
For example, the diameter of the second largest~\mbox{fragment} is expected to
be $0.76\,D_0$, its mass should be 0.435 the mass of the principal fragment,
and, if outgassing, its nongravitational acceleration should amount to 1.32
the acceleration of the principal fragment.  An expression similar to Equation
(14) can be derived for $D_i$ as a function of $D_{i+1}$ and ${\cal N}_{\rm
frg}(D_{\rm i+1})$.

The cross-sectional area of the nucleus' model by Szab\'o et al.\ (2012),
\mbox{$X_{\rm Sz} = 4300$ km$^2$}, is now to be interpreted as a sum of the
cross-sectional areas of all surviving fragments in the cluster that is
presumed to be optically~thin, \mbox{$X_{\rm frg} = X_{\rm Sz}$}; it serves
as another constraint on the distribution function,
\begin{equation}
X_{\rm frg} \!=\! {\textstyle \frac{5}{2}}D_0^{\frac{5}{2}} \!\!\!\!\:
 \int_{D_{\scriptstyle \star}}^{D_0} \!\!\! {\textstyle \frac{1}{4}}
 \pi D^2 \!\cdot\!\!\!\: D^{-\frac{7}{2}} dD \!=\! {\textstyle \frac{5}{4}} \pi
 D_0^2 \!\left( \!\!\sqrt{\frac{D_0}{D_{\textstyle \star}}} \!-\! 1 \!\!\right) \!,
\end{equation}
where $D_{\textstyle \star}$ is the diameter of the smallest fragment that
contributes to the observed cross-sectional area $X_{\rm frg}$.  As it follows
from Equation (15) that
\begin{equation}
D_{\textstyle \star} = D_0 \!\left(\!1\!+\!\frac{4X_{\rm frg}}{5\pi D_0^2}\!
 \right)^{\!\!-2}\!\!,
\end{equation}
the total number of fragments in the cluster is equal to
\begin{equation}
{\cal N}_{\rm frg}^{\textstyle \star} = {\cal N}_{\rm frg}(D_{\textstyle
 \star}) = \!\left(\! \frac{D_0}{D_{\textstyle \star}}\!\right)^{\!\!\frac{5}{2}}
 \!\! = \!\left( \! 1 \!+\! \frac{4X_{\rm frg}}{5 \pi D_0^2} \!\right)^{\!\!5}
\end{equation}
and an average fragment diameter in the distribution is
\begin{equation}
\langle D \rangle \!=\! \frac{{\displaystyle \int_{D_{\scriptstyle \star}}^{D_0}}\!
 \!\! D \!\cdot\!\!\!\:D^{-\frac{7}{2}} \, dD}{{\displaystyle \int_{D_{\scriptstyle
 \star}}^{D_0}} \!\!\! D^{-\frac{7}{2}} \, dD} \!=\! {\textstyle \frac{5}{3}}
 D_{\textstyle \star} \!\! \left[ 1 \!-\!\!\left( \! \frac{D_{\textstyle
 \star}}{D_0} \! \right)^{\!\!\frac{3}{2}} \right] \!\! \cdot \!\!\left[ 1
 \!-\!\! \left( \! \frac{D_{\textstyle \star}}{D_0} \!\right)^{\!\! \frac{5}{2}}
 \right]^{\!\!-1} \!\!\! .
\end{equation}
The mass contained in the cluster of fragments equals
\begin{eqnarray}
{\cal M}_{\rm frg} & = & {\textstyle \frac{5}{2}} D_0^{\frac{5}{2}} \!\!
 \int_{D_{\scriptstyle \star}}^{D_0} \!\!\! {\textstyle \frac{1}{6}} \pi \rho
 D^3 \!\cdot\!  D^{-\frac{7}{2}} \:\! dD \nonumber \\[-0.18cm]
 & & \\[-0.23cm]
 & = & \frac{5 \pi}{6} \rho D_0^3 \! \left( \!1 \!-\!
 \sqrt{\frac{D_{\textstyle \star}}{D_0}} \right) = 5 {\cal M}_0 \!
 \left[ 1 \!-\! ({\cal N}_{\rm frg}^{\textstyle \star})^{-\frac{1}{5}}
 \right]\!, \nonumber
\end{eqnarray}
where $\rho$ is a bulk density of the fragments and ${\cal M}_0$ the mass
of the principal fragment, \mbox{${\cal M}_0 \!=\! \frac{1}{6} \pi \rho D_0^3$},
{\vspace{-0.03cm}}which always exceeds 20\% of the cluster's total mass.  The
minimum and average fragment diameters, the number of fragments, and the
cluster's total mass, all derived from the observed cross-sectional area, are for
three principal fragment's diameters from the adopted range (Section~3) listed
in the top section of Table~1.{\vspace{-0.05cm}}
%
% Next, a critical distance, $s_{\rm frg}$, between the centers of fragments in
% the cluster is determined by a condition that no two neighboring fragments
% penetrate one another when at rest; it is given as an average of the diameters~of
% the principal fragment and second largest fragment,
%
% The critical diameter of a sphere occupied by the cluster of fragments in this
% setup is
%
% \begin{equation}
% D_{\rm frg}^\prime \!=\! \left( \!\frac{3\sqrt{2}}{\pi}{\cal N}_{\rm frg}^\star
%  \!\right)^{\!\!\frac{1}{3}} \!\! s_{\rm frg} \simeq 0.97 D_0 \! \left(
%  \!\frac{D_0}{D_\star} \!\right)^{\!\frac{5}{6}}
% \end{equation}
%
% and the cluster's critical spatial density
%
% \begin{eqnarray}
% \rho_{\rm frg}^\prime & = & \frac{20 \pi \sqrt{2}}{3} \! \left( \!1 \!+\!
%  2^{-\frac{2}{5}} \!\right)^{\!-3} \!\! \left( \! \frac{D_\star}{D_0}\!
%  \right)^{\!\!\frac{5}{2}}\!\!  \left( \! 1 \!-\! \sqrt{\frac{D_\star}{D_0}}
%  \right) \! \rho \nonumber \\[-0.28cm]
%  & & \\[-0.28cm]
%  & \simeq & 5.45 \! \left( \!\frac{D_\star}{D_0} \!\right)^{\!\!\frac{5}{2}}
%  \!\! \left( \!1 \!-\!\sqrt{\frac{D_\star}{D_0}} \right) \! \rho \,. \nonumber
% \end{eqnarray}
%

\section{Birth and Early Evolution of\\Compact Cluster}
The postulation of a breakup of the original nucleus of C/1995 O1 that was
triggered by Jupiter's tidal forces at or very near the time of closest approach
to the planet in November of $-$2251 (Section~1) is a plausible hypothesis for
the origin of the compact cluster, because if it had come into existence way
before the encounter, Jupiter's gravity field would have dissipated the cluster
into a long filament of fragments at the time of the event.

As described in detail in Paper 1, the comet has not approached Jupiter to within
0.7~AU since the encounter, but it passed through perihelion about 13~months
\mbox{after} the near miss, at a heliocentric distance almost identical to that
in 1997.  During the more than 42 centuries the size distribution of fragments,
especially near its lower end should have changed dramatically, as illustrated
by Hamilton \& Burns (1992) for asteroids. All fragments smaller than a few
centimeters across were removed by radiation pressure within years after the
breakup.  Of the remaining fragments, including the more sizable ones, all those
moving in prograde orbits were soon lost as well.\footnote{Fragments larger than
a few centimeters but smaller than $D_{\textstyle \star}$ in diameter (Table~1)
were probably removed by the outgassing-driven momentum, whose influence --- at
least at heliocentric distances of up to a few AU --- resembles that of solar
radiation pressure, varying as an inverse diameter.  For example, scaling the
observed effect, a fragment 50~m in diameter is expected at 1~AU from the Sun to
be subjected to an outgassing-driven acceleration on the order of 0.002 the Sun's
gravitational acceleration, equivalent to a radiation-pressure acceleration on
a particle of about 1.5~mm in diameter at the same density.  As activity ceases
at larger heliocentric distances, so does the momentum and, accordingly, the
resulting orbit-integrated effect is more heavily dependent on its magnitude
near the Sun than is radiation pressure.} For fragments in retrograde orbits
the situation is less clear, but, as a rule of thumb, one can assume that at
least 50\% of them failed to survive until the 1997 apparition.  Because the
initial orbital velocities of fragments relative to the cluster's center of
mass (dictated by the rotation of the original nucleus at the time of breakup)
were very low, the high collisional rate (Section~8) must have soon brought a
near equilibrium between the numbers of fragments moving in prograde and
retrograde orbits, so for the sake of argument one can assume that about one
quarter of all fragments with diameters larger than $D_{\textstyle \star}$
survived until the 1997 apparition.

If no fragment with a diameter smaller than $D_{\textstyle \star}$ but one
quarter of the fragments with diameters \mbox{$D_{\textstyle \star} \!\leq\!
D \!\leq\! D_0$} survived to the 1997 apparition, the mass ${\cal M}_{\rm C}$
of the original nucleus of C/1995~O1 at the time of encounter with Jupiter
is estimated to have equaled
\begin{equation}
{\cal M}_{\rm C} = {\cal M}_{\rm C} \sqrt{D_{\textstyle \star}/D_0} + 4
 {\cal M}_{\rm frg},
\end{equation}
where the mass ${\cal M}_{\rm frg}$ is given by Equation~(19).  Solving
Equation~(20) for ${\cal M}_{\rm C}$, one finds in terms of the principal
fragment's parameters:
\begin{equation}
{\cal M}_{\rm C} = \frac{10\pi}{3} \rho D_0^3 = 20 {\cal M}_0.
\end{equation}

This is of course a very crude estimate, but it is needed only for an
assessment of the probability of tidal breakup in Section~7.  In the range
of the three solutions in Table~1, the original nucleus was between 2.1 and
7.3$\,\times$10$^{18}$g in mass and between 21 and 33~km{\vspace{-0.04cm}}
in diameter (at the adoped bulk density of 0.4~g~cm$^{-3}$), a considerably
less impressive and statistically more probable size than 74~km.  Yet, because
of the greater mass and the very limited volume of space involved, the
collisional rate in the early phase of cluster evolution should have been
orders of magnitude higher than at the 1997 apparatition.

Judging from the numerical experiments by Hamilton \& Burns (1991, 1992),
much of the dust generated~in~the course of fragmentation is expected to have
been blown away by solar radiation pressure, especially near the perihelion
passage 13 months after the Jovian encounter, whereas smaller, active
boulder-sized objects were subjected to nongravitational accelerations
whose effects ultimately were not unlike those of radiation pressure.
I return to the issue of small fragments in Section~8, after introducing a
constraint on the collisional rate.

\begin{table}[b]
\vspace{-3.57cm}
\hspace{4.22cm}
\centerline{
\scalebox{1}{
\includegraphics{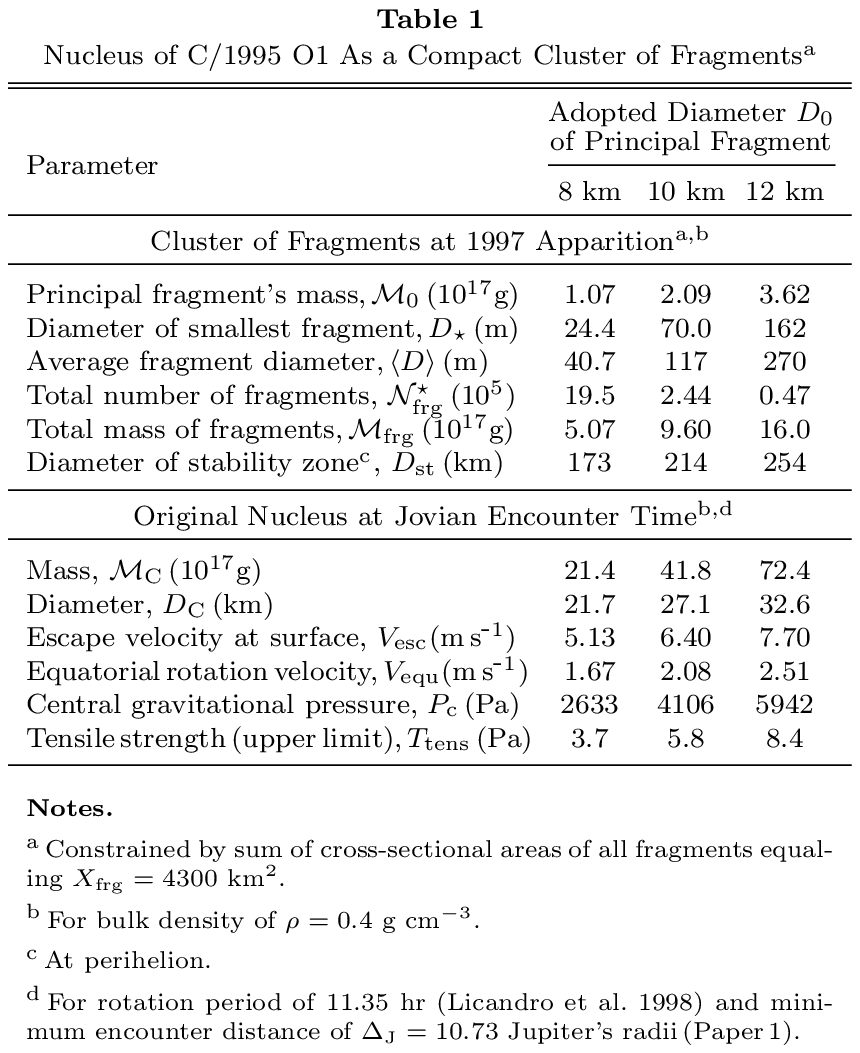}}} %  From t1_HB2.tex
\vspace{-15.17cm}
\end{table}

\section{Tidal Force of Jupiter vs Dimensions and\\Tensile Strength of 
 Original Nucleus}
Consider the original, pre-encounter nucleus of comet C/1995 O1 as a porous rigid
rotating spherical aggregate of dust and ices in static equilibrium, moving in
a strongly hyperbolic orbit about Jupiter and exposed to its tidal forces.  A
comprehensive stress theory by Aggarwal \& Oberbeck (1974) suggests that at a
Jovicentric distance $\Delta_{\rm J}$ fissures should start propagating from
the equatorial regions on the nucleus' surface at which Jupiter is rising
above, or setting below, the local horizon, when the tensile strength, $T_{\rm
tens}$, satisfies a condition
\begin{equation}
T_{\rm tens} = {\textstyle \frac{10}{19}} P_{\rm c} \!\left(\! \frac{R_{\rm
 J}}{\Delta_{\rm J}} \! \right)^{\!\!3} \! \frac{\rho_{\rm J}}{\rho},
\end{equation}
where $R_{\rm J}$ and $\rho_{\rm J}$ are Jupiter's radius and mean density,
$\rho$ is again the nucleus' bulk density, and
\begin{equation}
P_{\rm c} = {\textstyle \frac{1}{6}} \pi G \rho^2 D_{\rm C}^2
\end{equation}
is the gravitational pressure in the center of the nucleus of diameter
$D_{\rm C}$, with $G$ being~the~gravitation\-al constant.  Inserting
\mbox{$\Delta_{\rm J}/R_{\rm J} = 10.73$} (Paper~1), \mbox{$\rho_{\rm J} =
1.33$ g cm$^{-3}$}, and \mbox{$\rho \!=\! 0.4$ g cm$^{-3}$}, the tensile strength
is between~3.7~Pa and 8.4~Pa for the original nucleus 21.7~km to 32.6~km in
diameter (Table~1).  These tensile strength values are near the lower end of a
range reported by Groussin et~al.\ (2015) [3~Pa] and Basilevsky et al.\ (2016)
[$>$1.5~Pa]~from their studies of outcropped consolidated material in cliff-like
features on the nucleus' surface of comet 67P. As~a short-period comet, 67P was
exposed to processes~such~as sintering, which have a tendency to increase the
strength of material and which C/1995~O1 is not expected~to~have experienced
before the encounter.  Its tidal breakup in close proximity of Jupiter should
accordingly be judged as plausible.

Once a fissure began to propagate inside the comet's nucleus, a separation of
the early fragments was only a matter of time.  The fragmentation may have been
assisted by the rotational velocity, which --- if the spin rate was close to that
observed during the 1997 apparition (a rotation period of $\sim$11.35~hr; Licandro
{\vspace{-0.04cm}}et al.\ 1998) --- amounted to between 1.7~m~s$^{-1}$ and~2.5~m~s$
^{-1}$. Since the velocity of escape equaled~5.1~m~s$^{-1}$~to~7.7~m~s$^{-1}$, the
early fragments moved along ballistic trajectories, resulting in imminent impacts,
further fragmentation, and random walk of secondary fragments superposed on their
rotationally-driven motions.~The~momentum~should~have progressively built up to
make the developing \mbox{cluster}~of colliding fragments slowly expand around
its center~of mass.  On the assumption that the cluster was~\mbox{gradually}
acquiring spherical symmetry,{\vspace{-0.08cm}} a root-mean-squared cir\-cular
velocity, \mbox{$\langle V_{{\rm circ},\ell}^2 \rangle{\mbox{\raisebox{0.2ex}{$^{\!
{\frac{1}{2}}\!}$}}}$}, at distances between \mbox{$\ell \!-\! d\ell$} and~$\ell$
should satisfy a condition
\begin{equation}
\langle V_{{\rm circ},\ell}^2 \rangle \, 4\pi\ell^2 d\ell = \frac{G{\cal
 M}_{\rm f}(\ell)}{\ell} \, 4\pi \ell^2 d\ell,
\end{equation}
where ${\cal M}_{\rm f}(\ell)$ is the mass of the fragments located at distances
smaller than $\ell$ from the center of mass, to whose gravitational attraction
the fragments orbiting between \mbox{$\ell \!-\! d\ell$} and $\ell$ are subjected
to.  All fragments at distances greater than $\ell$ represent minor perturbers.
Assuming the cluster's spatial density to be independent of the distance from
the center of mass, ${\cal M}_{\rm f}(\ell)$ varies as the volume confined within
$\ell$, and if $D_{\rm frg}$~is the cluster's diameter, the{\vspace{-0.06cm}}
root-mean-squared circular velocity averaged over the cluster, $\langle V_{\rm
circ}^2 \rangle {\mbox{\raisebox{-0.6ex}{$^{\!^{\frac{1}{2}}\!}$}} }$, is
determined by a condition{\vspace{-0.05cm}}
\begin{equation}
\langle V_{\rm circ}^2 \rangle \!\! \int_{0}^{\frac{1}{2}D_{\rm frg}} \!\!
 \ell^{\:\!2}
 d\ell = \!\! \int_{0}^{\frac{1}{2}D_{\rm frg}} \frac{G {\cal M}_{\rm frg}}{\ell}
 \! \left( \! \frac{2 \ell}{D_{\rm frg}} \! \right)^{\!\!3} \! \ell^{\:\!2} d\ell,
\end{equation}
from which
\begin{equation}
\langle V_{\rm circ}^2 \rangle^{\!\frac{1}{2}} = \left( \!\frac{6 \:\! G
 {\cal M}_{\rm frg}}{5D_{\rm frg}} \! \right)^{\!\!\frac{1}{2}}\! , 
\end{equation}
where ${\cal M}_{\rm frg}$ is given by Equation~(19).

\section{Rate of Collisions Among Fragments and\\the Cluster's Size}
Consider a spherical fragment of a diameter $D_i$ moving with a velocity $V_{i,j}$
relative to another fragment whose diameter is $D_j$.  The cross-sectional area for
a collison~between these two fragments equals (e.g., Kessler 1981){\vspace{-0.05cm}}
\begin{equation}
\sigma_{i,j} = \frac{\pi}{4} (D_i \!+\! D_j)^2 ,
\end{equation}
where the fragments' escape velocity is being neglected.  If $\nu$ is
a number density of fragments in the cluster, i.e., their number per
unit volume, an average number of~collisions {\vspace{-0.07cm}}that the
fragment with a diameter $D_i$ experiences per unit time,\,$\dot{N}_{\rm
coll}^{(i)}$, then equals{\vspace{-0.05cm}}
\begin{equation}
\dot{N}_{\rm coll}^{(i)} = \nu \langle \sigma_i \rangle \langle V_{{\rm rel},i}^2
 \rangle^{\!\frac{1}{2}},
\end{equation}
where $\langle \sigma_i \rangle$ is an average collisional{\vspace{-0.06cm}}
cross-sectional area for a fragment of diameter $D_i$ and{\vspace{-0.04cm}}
$\langle V_{{\rm rel},i}^2 \rangle^{\!\frac{1}{2}}$ is its root-mean-squared
impact velocity averaged over all fragments with which it collides.  If the
population of fragments has the size distribution introduced in Section~4,
the collisional cross-sectional area $\langle \sigma_i \rangle$ equals
\begin{eqnarray}
\langle \sigma_i \rangle & = & - {\textstyle \frac{5}{2}} D_{\textstyle
 \star}^{\frac{5}{2}} \!\! \left[1 \!-\! \left(\!\frac{D_{\textstyle \star}}{D_0}
 \! \right)^{\!\!\frac{5}{2}} \right]^{\!-1} \!\!\!\! \int_{D_{\scriptstyle
 \star}}^{D_0} \!\!\! \sigma_{i,j} D_j^{-\frac{7}{2}} dD_j \nonumber \\[-0.35cm]
 & & \\[-0.05cm]
 & = & \frac{\pi}{4} D_i^2 \!\left[1 \!+\! \frac{5}{\Gamma_5} \frac{D_{\textstyle
 \star}}{D_i} \!\left(\! \frac{2\Gamma_3}{3} \!+\! \frac{D_{\textstyle \star}}{D_i}
 \!\right) \! \right] \!, \nonumber
\end{eqnarray}
where{\vspace{-0.11cm}}
\begin{equation}
\Gamma_m = \!\sum_{k = 0}^{m-1} \!\left(\! \frac{D_{\textstyle \star}}{D_0}
 \!\right)^{\!\!\frac{1}{2}k} \!\! = \frac{1 \!-\! \sqrt{(D_{\textstyle
 \star}/D_0)^{m}}}{1 \!-\! \sqrt{D_{\textstyle \star}/D_0}} .
\end{equation}
Averaging now $D_i$ over all fragment diameters between $D_{\textstyle \star}$
and $D_0$, the mean cross-sectional area $\langle \sigma \rangle$ for collisions
between any two such fragments becomes
\begin{eqnarray}
\langle \sigma \rangle & = & - {\textstyle \frac{5}{2}} D_{\textstyle
 \star}^{\frac{5}{2}} \!\!\left[1 \!-\! \left(\! \frac{D_{\textstyle
 \star}}{D_0} \!\right)^{\!\!\frac{5}{2}} \right]^{\!-1} \!\!\!\!
 \int_{D_{\scriptstyle \star}}^{D_0} \!\!\! \langle \sigma_i \rangle
 D_i^{-\frac{7}{2}} dD_i \nonumber \\[-0.35cm]
 & & \\[-0.05cm]
 & = & \frac{5 \pi}{2 \Gamma_5} D_{\textstyle \star}^2 \!\left( \!1 \!+\!
 \frac{5 \Gamma_3^2}{9 \Gamma_5} \:\!\!\right) \!, \nonumber
\end{eqnarray}
and an average number of collisions per unit time, $\dot{N}_{\rm coll}$,
experienced by the fragments with diameters between $D_{\textstyle \star}$ and
$D_0$ is
\begin{equation}
\dot{N}_{\rm coll} = \nu \langle \sigma \rangle \langle V_{\rm rel}^2
 \rangle^{\!\frac{1}{2}},
\end{equation}
where $\langle V_{\rm rel}^2 \rangle^{\!\frac{1}{2}}$ is their root-mean-squared
velocity averaged over the cluster.  I now assume that this velocity varies in
proportion to the root-mean-squared average circular velocity, derived in
Section~5,
\begin{equation}
\langle V_{\rm rel}^2 \rangle^{\!\frac{1}{2}} = \eta \langle V_{\rm circ}^2
 \rangle^{\!\frac{1}{2}}.
\end{equation}
In his elaborate collisional model for the asteroid population, Dohnanyi (1969)
adopted an impact velocity equivalent to \mbox{$\eta \simeq 0.29$}.  However,
the motions of asteroids in the belt are much more organized than are fragments
in the proposed cluster, for which $\eta$ should be much greater but not
exceeding unity, because there is not enough energy in the system to achieve
\mbox{$\langle V_{\rm rel}^2 \rangle \gg \langle V_{\rm circ}^2 \rangle$}. 

An important constraint follows from Schr\"{a}pler et al.'s (2012) microgravity
experiments that showed that in order for fluffy dust aggregates to fragment upon
impact, their relative velocity should be at least 0.4~m~s$^{-1}$, since at lower
velocities they bounce or stick.  Similar independent experiments by Gunkelmann
et al.\ (2016) suggest that for highly porous submicron-grain~\mbox{agglomerates},
comparable in porosity to cometary nuclei, the minimum impact velocity triggering
fragmentation is still lower, at $\sim$0.17~m~s$^{-1}$.  Since the impact velocity
is a function of the cluster size, which in turn depends on the velocity, the
parameter $\eta$ is constrained but not well determined.

\begin{figure*}[t]
\vspace{-4.4cm}
\hspace{-0.67cm}
\centerline{
\scalebox{0.85}{ % 0.82
\includegraphics{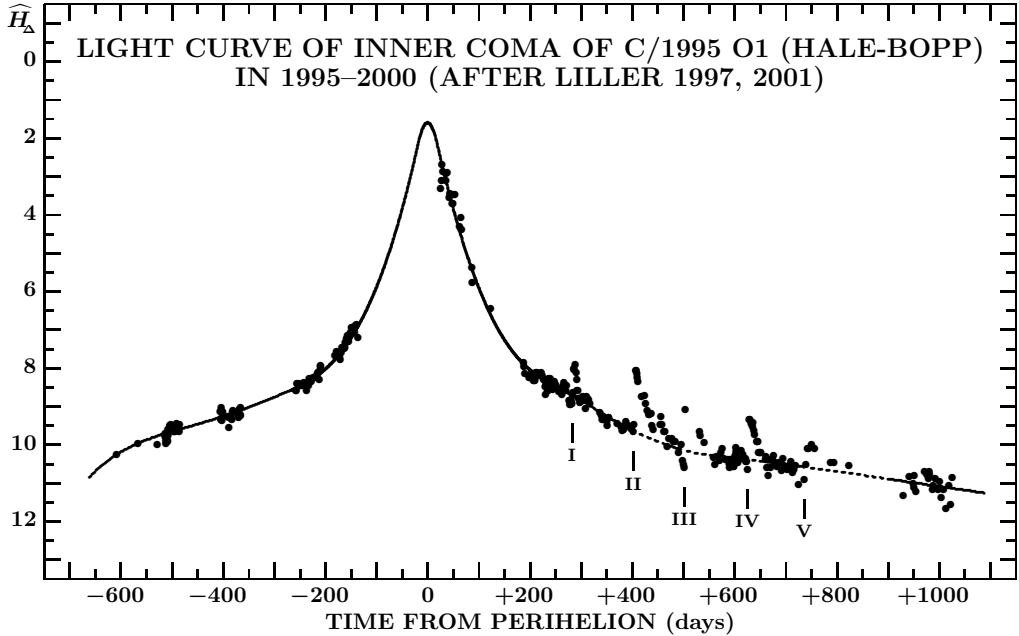}}} %  From: f1_HB2.tex, originally obr33_tichy-17P.tex
\vspace{-12.39cm}
\caption{Visual light curve of the inner coma of comet C/1995 O1 (normalized to
a nucleus-centered field of 24\,660~km on a side when observed from a distance
of 1~AU) based on the CCD observations by Liller (1997, 2001) made with a 20-cm
f/1.5 Celestron camera between 1995 August 2 and 2000 January 21 (608 days
preperihelion to 1025 days post-perihelion).  The curve is very smooth before
perihelion, but dotted with at least five prominent flare-ups after perihelion,
between 1998 January and 1999 April, when the comet was 4 AU to 8.2 AU from the
Sun.  The onset times of the events are marked \mbox{I--V}.  Their amplitudes,
\mbox{0.7--1.6 mag}, imply a sudden increase in the cross-sectional area of
the dust ejecta in the 34$^{\prime\prime}$ field of up to nearly 3 million
km$^2$ at an assumed geometric albedo of 0.04 (Table~2).{\vspace{0.5cm}}}
\end{figure*}

With the number density of fragments in the cluster being{\vspace{-0.2cm}}
\begin{equation}
\nu = \frac{6 {\cal N}_{\rm frg}^{\textstyle \star}}{\pi D_{\rm frg}^3} = 
 \frac{6}{\pi D_{\rm frg}^3} \!\left( \! \frac{D_0}{D_{\textstyle \star}}
 \!\right)^{\!\!\frac{5}{2}}\!\!,
\end{equation}
I insert from Equations (31), (33), and (34) as well as from (26), (19), and
(15) into Equation~(32) and obtain for a mean free time{\vspace{-0.04cm}}
between two consecutive collisions, \mbox{$\tau_{\rm coll} \!=\! \dot{N}_{\rm
coll}^{-1}$}, an expression
\begin{equation}
\tau_{\rm coll} = \frac{\sqrt{5}}{30\eta}\, \frac{\Gamma_5^2}{\Gamma_5 \!+\!
 {\textstyle \frac{5}{9}}\Gamma_3^2} \, (G \rho X_{\rm frg})^{-\frac{1}{2}}
 D_{\textstyle \star}^{\frac{1}{4}} D_0^{-\frac{11}{4}} D_{\rm frg}^{\frac{7}{2}}.
\end{equation}
The mean free time depends heavily on the dimensions of the cluster and the
principal fragment, but only weakly on the size of the smallest fragment,
which besides its quartic root enters the expression via the sums
$\Gamma_3$~and~$\Gamma_5$.

Equation (35) can in principle serve to determine the size of the cluster at
the 1997 apparition as a function of $D_0$, once the collisional mean free time
$\tau_{\rm coll}$ and the impact velocity parameter $\eta$ are known.  Although
the exact dimensions of the cluster at the 1997 apparition are unknown, they are
rather strongly constrained; the cluster's gravitational stability over more
than four millennia requires that its diameter not exceed the stability limit
at large heliocentric distances, where the comet spends nearly all of its life.
Another upper limit is provided by the HST images taken between October 1995 and
October 1996; one pixel, which the cluster's diameter should never exceed by
more than a factor of about two, equaled between 90~km and 220~km. On the other
hand, assuming that the cluster was optically thin, its diameter should much
exceed Szab\'o et al.'s (2012) 74~km.

\section{Liller's Detection of Recurring Flare-Ups and Their Proposed
 Interpretation}

A pair of important papers on C/1995 O1 was written by Liller (1997, 2001).
He monitored the brightness of the inner coma using a 20-cm f/1.5 Celestron
camera, a CCD detector, and a filter to obtain magnitudes in the $V$ system.
His dataset consists of exposures on 360 nights, {\vspace{-0.02cm}}covering
a time period of nearly 4$\frac{1}{2}$~yr, from 1995 August 2 (608 days before
perihelion) to 2000 January 21 (1025 days after perihelion).  The comet was
7.06~AU and 10.28~AU from the Sun on, respectively, the former and the latter
dates.  {\vspace{-0.06cm}}For each exposure Liller measured an apparent magnitude
$\widehat{H}$ in a square field of 34$^{\prime\prime}$ on a{\vspace{-0.07cm}}
side, with the nucleus in its center, and converted it to $\widehat{H}_\Delta$,
by removing the effect of a variable field size due to the comet's changing
geocentric{\vspace{-0.07cm}} distance $\Delta$, with an expression
\mbox{$\widehat{H}_\Delta\!=\!\widehat{H} \!-\!  2.5 \log \Delta$}.  This
light curve is reproduced in Figure~1 as a function of the time from perihelion.  

Liller (2001) called attention to a prominent anomaly in the light curve, which
is very smooth before perihelion but dotted with at least five flare-ups after
perihelion.  He noted that the flare-ups were distributed approximately uniformly
in time, with gaps of 96~days~to~125~days; that the peak amplitudes ranged from
0.7~mag to 1.6~mag; that their heliocentric distances varied from 4.0~AU to 8.1~AU;
and that the expansion velocities{\vspace{-0.01cm}} of the ejected material were
confined to a range from 62~m~s$^{-1}$ to 217~m~s$^{-1}$.  The peaks of the five
flare-ups were observed between 1998 January 11 and 1999 April 14, and there could
well have been two additional flare-ups, one in late August 1997 and the other in
mid-October 1999.  Liller expressed his belief that the flare-ups were caused by
the nucleus' recurring activity, not by collisions with asteroids, but he did not
propose any specific active process.

\begin{table*}
\vspace{-4.08cm}
\hspace{-0.54cm}
\centerline{
\scalebox{1}{
\includegraphics{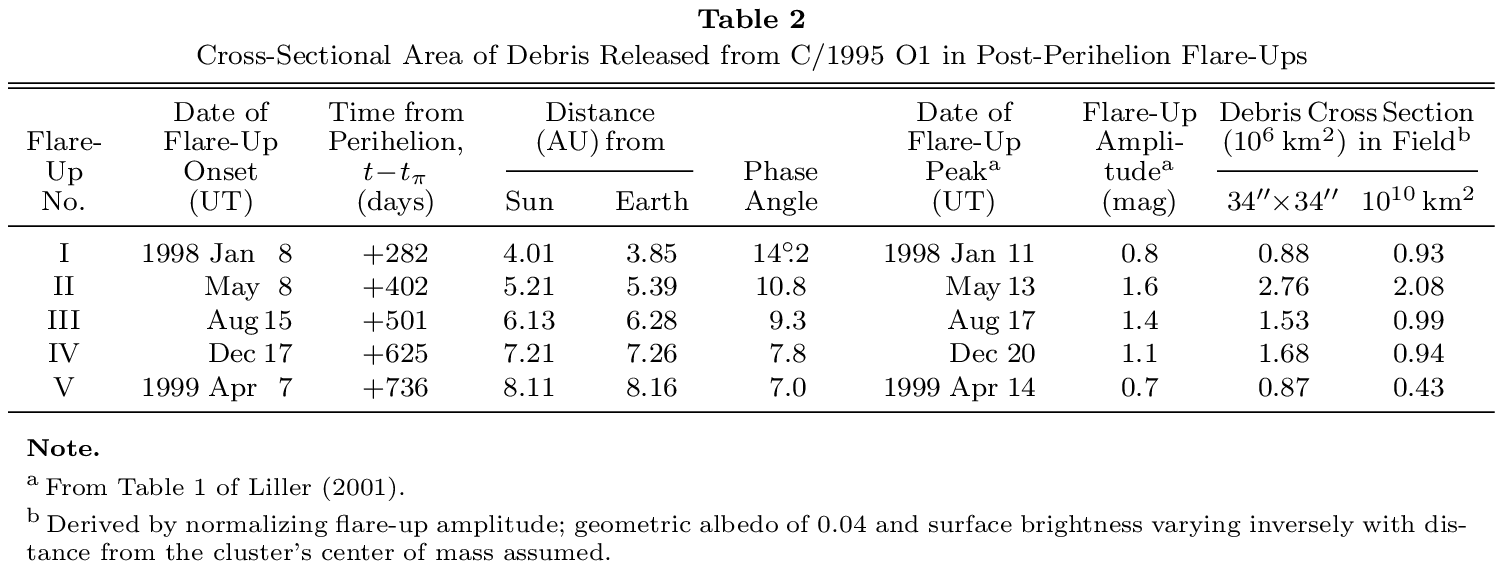}}} %  From t2_HB2.tex
\vspace{-19.24cm} % -19.35
\end{table*}

Because the flare-ups were observed at large heliocentric distances, they could
not be triggered by a suddenly elevated sublimation rate of water ice.  Instead,
the driver would have to have been explosions of carbon monoxide (possibly
assisted by carbon dioxide).  Unfortunately, as seen from Figure~9 of Paper~1,
there were no obvious peaks on the carbon-monoxide production curve temporally
coinciding with Liller's flare-up times in Figure~1.  The single strongly
elevated carbon-monoxide production rate, based on an observation made with
an instrument onboard the Infrared Space Observatory (ISO) on 1998 April 6
(370~days after perihelion, when the comet was 4.9~AU from the Sun) precedes
the flare-up II in Figure~1 by five weeks, and its origin is unclear.  Thus,
an intrinsic, carbon-monoxide driven event was not a primary trigger of the
five (or six) post-perihelion flare-ups detected by Liller (2001).

Could each of the observed flare-ups contain~in~fact~the debris of particular
fragments in the cluster that collided with one another and broke up?  In order
to examine the implications of this hypothesis, let the masses of~the colliding
fragments be ${\cal M}_i$ and ${\cal M}_j$, respectively, and let their relative
velocity $(V_{\rm rel})_{i,j}$ upon impact be high enough so that they both
fracture (rather than bounce or stick together).  In addition, let the slope
of the size distribution of the debris generated by this collision equal the
slope of the size distribution of the cluster's fragments, in which case the
mass contained in the collisional debris\\[-0.2cm]
\begin{equation}
({\cal M}_{\rm deb})_{i,j} = {\cal M}_i + {\cal M}_j,
\end{equation}
amounts to, in analogy to Equation (19),
\begin{eqnarray}
({\cal M}_{\rm deb})_{i,j} & = & \frac{5 \pi}{6} \rho \, (D_{\rm max})_{i,j}^3
 \!\left[ 1 \!-\! \sqrt{\frac{(D_{\rm min})_{i,j}}{(D_{\rm max})_{i,j}}} \,
 \right] \nonumber \\[-0.2cm]
& & \\[-0.2cm]
& = & 5\,({\cal M}_{\rm max})_{i,j} \!\left[ 1 \!-\! \sqrt{\frac{(D_{\rm
 min})_{i,j}}{(D_{\rm max})_{i,j}}} \,\right] \!, \nonumber
\end{eqnarray}
where\,$(D_{\rm max})_{i,j}$\,and\,$(D_{\rm min})_{i,j}$\,are, respectively, the
diam\-eters of the largest and smallest pieces in the debris{\vspace{-0.02cm}}
of colliding {\vspace{-0.05cm}}fragments $i$ and $j$, while \mbox{$({\cal M}_{\rm
max})_{i,j} \!=\! \frac{1}{6} \pi \:\!\!\rho \:\!(D_{\rm max})_{i,j}^3$} is the
mass of the largest piece.  Furthermore, similarly~to Equation~(15), the
cross-sectional area of the debris is
\begin{equation}
(X_{\rm deb})_{i,j} = \frac{5 \pi}{4} (D_{\rm max})_{i,j}^2 \!\left[
 \sqrt{\frac{(D_{\rm max})_{i,j}}{(D_{\rm min})_{i,j}}} \!-\! 1 \right]
\end{equation}
and, following Equation (17), the number of pieces in the debris field becomes
\begin{equation}
(N_{\rm deb})_{i,j}^{\textstyle \star} = \left[ \frac{(D_{\rm
 max})_{i,j}}{(D_{\rm min})_{i,j}} \right]^{\!\frac{5}{2}} \! = \left[ 1 \!+\!
 \frac{4\:\! (X_{\rm deb})_{i,j}}{5 \pi (D_{\rm max})_{i,j}^2} \right]^{\!5} \!\!.
\end{equation}

The five flare-ups reported by Liller (2001) are summarized in Table~2, the
first eight columns of which are self-explanatory.  Dropping the subscripts
$i$ and $j$, the critical quantity is the cross-sectional area $X_{\rm obs}$,
computed from a normalized amplitude of the flare-up's light curve by assuming
a geometric albedo of 0.04 and allowing for the phase effect with the use of
the Marcus (2007) modification~of~the~Henyey-Greenstein scattering law.
Observed in a \mbox{34$^{\prime\prime} \!\! \times \!  34^{\prime\prime}$}
aperture, $X_{\rm obs}$ is listed in the penultimate column.  In the last
column it is scaled to a constant linear aperture of 10$^5$\,km on a side,
assuming that the surface brightness of the flare-up varied inversely as the
distance from the cluster's center of light.
  
%
% Once $X_{\rm obs}$ is determined, the observed mean
% collision rate --- 3.2 per year (the mean free time of 0.31~yr) if there were
% five flare-ups, or 4.0 per year (the mean free time of 0.25~yr) if six ---
% should one allow to iterate Equation~(35) to find the diameter of the cluster
% as a function of the velocity parameter $\eta$.
%

The tabulated cross-sectional data appear to be in a range of \mbox{$\sim$1--3
million km$^2$} in the 34$^{\prime\prime\!}$ by~34$^{\prime\prime}$ field,
more than two orders of magnitude greater than the cross-sectional area of
the comet's nucleus, based on the work by Szab\'o et al.\ (2012) and
identified here with the total cross section of the proposed cluster of
fragments.  Some of this disparity is explained by the presence of optically
effective, very fine dust in the flare-up debris, in contrast to the absence
of fragments less than a few tens of meters across in the cluster once a
flare-up fades away (Table~1).  Accordingly, I adopt \mbox{$D_{\rm min}
\simeq 0.3 \; \mu$m} in Equations~(37) to (39), meaning it refers to the
submicron-sized grains embedded in porous aggregate particles (e.g.,
Brownlee 1985)\footnote{Aggregate particles of this kind were recently
reported to make up major part of the dust population of comet
67P/Churyumov-Gerasimenko (Della Corte et al.\ 2015; Merouane et al.\
2016).{\vspace{-0.33cm}}} that are believed to make up much of the
refractory mass of the fragments.

Another part of the disparity derives from the fact that a substantial increase
in the surface area during the collisional fragmentation triggered off an
increased production of carbon monoxide from the newly exposed surface, which
necessarily entailed an increased production of dust.  Evidence of CO-driven
dust is implied by Liller's (2001) remark that the expansion{\vspace{-0.04cm}}
rate of ejecta during the flare-ups exceeded 60~m~s$^{-1}$, about two orders
of magnitude higher than the impact velocity.  This means that a flare-up's
amplitude and the corresponding cross-sectional area $X_{\rm obs}$ consisted
of at least two different components, $X_{\rm deb}$ referring only to the
low-velocity mass.  Since, as already noted, no spike was apparent in the
carbon-monoxide production rate at the times of the flare-ups, the amplitude
of the high-velocity component did not exceed the overall scatter in the CO
production rate, which according to Table~19 of Paper~1 amounted to 10$^{\pm 0.12}$,
translating to a peak amplitude of 0.60~mag.  Table~2 shows that Liller was
able to detect flare-ups with an amplitude of 0.7~mag, so that a conservative
estimate for a minimum detectable amplitude of the low-velocity component is
$\sim$0.1~mag, equivalent to a lower limit of the cross-sectional area of
collisional debris of \mbox{$X_{\rm lim} \simeq 1 \times$10$^5$\,km$^2$},
which can now readily be equated with $X_{\rm deb}$ from Equation~(38).

The issue now is how does $X_{\rm lim}$ compare with the cross-sectional area
of the debris generated by a collision of two least massive fragments in the
cluster, of a diameter $D_{\textstyle \star}$.  The mass of this debris is from
Equation~(36),
\begin{equation}
{\cal M}_{\rm deb}(D_{\textstyle \star},D_{\textstyle \star}) =
 2{\cal M}(D_{\textstyle \star}) = \frac{\pi}{3} \rho D_{\textstyle \star}^3
\end{equation}
and, on the other hand, since \mbox{$D_{\rm min} \!\ll\! D_{\textstyle \star}$},
in terms of the largest piece of the debris, from Equation (37),
\begin{equation}
{\cal M}_{\rm deb}(D_{\textstyle \star},D_{\textstyle \star}) = \frac{5\pi}{6}
 \rho D_{\rm max}^3.
\end{equation}
Using these equations to eliminate $D_{\rm max}$ and given again that \mbox{$D_{\rm
min} \!\ll\! D_{\textstyle \star}$}, one obtains for the cross-sectional area of
the debris from Equation~(38)\\[-0.15cm]
\begin{equation}
X_{\rm deb}(D_{\textstyle \star},D_{\textstyle \star}) = \!\left(\!  \frac{5}{128}
 \!\right)^{\!\!\frac{1}{6}} \!\!\pi D_{\textstyle \star}^{\frac{5}{2}}
 D_{\rm min}^{-\frac{1}{2}}.
\end{equation}
For the three scenarios from Table~1, the cross-sectional areas of
the debris generated by a collision of the smallest fragments in the cluster
are 10~km$^2$ for \mbox{$D_0 = 8$ km}, 137~km$^2$ for \mbox{$D_0 = 10$ km}, and
1116~km$^2$ for \mbox{$D_0 = 12$ km}.  These cross sections are all smaller
than $X_{\rm lim}$, which indicates that Liller (2001) missed these collisions
because the triggered flare-ups had amplitudes that were too shallow to detect.
Accordingly, Equation (35) needs to be corrected for incomplete statistics before
the mean collisional rate (or the mean free time between collisions) based on
Liller's flare-up observations can be employed to derive the dimensions of the
cluster.  A correction is to be applied in such a way that the smaller of any
pair of colliding fragments should be allowed to have a diameter from the entire
range of \mbox{$D_{\textstyle \star} \!\leq\! D \!<\!  D_0$}, whereas the larger
one only from a range of \mbox{$D_{\rm lim} \!\leq\! D \!\leq D_0$}; the task is
to find $D_{\rm lim}$ such that a collision involving this fragment generates
{\vspace{-0.06cm}}a debris whose cross-sectional area equals~$X_{\rm lim}$
(Liller's detection limit); all collisions with a rate of $\dot{N}_{\rm
coll}^{\textstyle \star}$, for which the diameter of the larger fragment is
from a range \mbox{$D_{\textstyle \star} \!\leq\! D \!<\! D_{\rm lim}$}, are
to be excluded from the count.

The procedure is very similar to the one used in Equations (40) to (42),
starting now with a condition
\begin{eqnarray}
\!\!\!\!\! {\cal M}_{\rm deb}(D_{\rm lim},D_{\textstyle \star}) & =
 & {\cal M}(D_{\rm lim}) + {\cal M}(D_{\textstyle \star}) \nonumber \\[-0.1cm]
 & & \\[-0.28cm]
 & = & \frac{\pi}{6} \rho D_{\rm lim}^3 \!\left[ 1 \!+\!\! \left(\!
 \frac{D_{\textstyle \star}}{D_{\rm lim}} \!\right)^{\!\!3} \right]
 \!\simeq\! \frac{\pi}{6} \rho D_{\rm lim}^3 \nonumber
\end{eqnarray}
and resulting in
\begin{equation}
D_{\rm lim} = 5^{-\frac{1}{15}} \!\!\left(\! \frac{4X_{\rm lim}}{\pi}
 \!\right)^{\!\!  \frac{2}{5}} \!\! D_{\rm min}^{\frac{1}{5}} = 1.2\;{\rm km}.
\end{equation}
The collisional rate for Liller's observations becomes
\begin{equation}
(\dot{N}_{\rm coll})_{\rm obs} = \dot{N}_{\rm coll} \!-\! \dot{N}_{\rm
 coll}^{\textstyle \star},
\end{equation}
where $\dot{N}_{\rm coll}$ is given by Equation~(32) and the respective
mean free time between collisions by Equation~(35).{\vspace{-0.07cm}}
The collisional rate $\dot{N}_{\rm coll}^{\textstyle \star}$ is similarly
expressed as
\begin{equation}
\dot{N}_{\rm coll}^{\textstyle \star} = \nu^{\textstyle \star} \langle
 \sigma^{\textstyle \star} \rangle \langle V_{\rm rel}^2 \rangle^{\frac{1}{2}}
\end{equation}
where
\begin{eqnarray}
\nu^{\textstyle \star} & = & \frac{6}{\pi D_{\rm frg}^3} \!\left[\!\left(\!
 \frac{D_0}{D_{\textstyle \star}} \!\right)^{\!\!\frac{5}{2}} \!\!-\!\left(\!
 \frac{D_0}{D_{\rm lim}} \!\right)^{\!\!\frac{5}{2}} \right] \nonumber \\[-0.05cm]
 & & \\[-0.35cm]
 & = & \frac{6 \Psi_5}{\pi D_{\rm frg}^3} \!\left(\! \frac{D_0}{D_{\textstyle
 \star}} \!\right)^{\!\!\frac{5}{2}} \!\!\!\cdot \!\!\!\:\left(\! 1 \!-\!
 \sqrt{\frac{D_{\textstyle \star}}{D_{\rm lim}}} \right) \nonumber
\end{eqnarray}
and, in analogy to Equation (31),
\begin{equation}
\langle \sigma^{\textstyle \star} \rangle = \frac{5\pi}{2 \Psi_5} D_{\textstyle
 \star}^2 \!\left(\! 1 \!+\! \frac{5 \Psi_3^2}{9 \Psi_5} \right) \!,
\end{equation}
with
\begin{equation}
\Psi_m = \!\sum_{k=0}^{m-1} \!\left(\! \frac{D_{\textstyle \star}}{D_{\rm lim}}
 \!\right)^{\!\!\frac{1}{2}k} \!\! = \frac{1 \!-\!
 \sqrt{(D_{\textstyle \star}/D_{\rm lim})^m}}{1 \!-\! \sqrt{D_{\textstyle
 \star}/D_{\rm lim}}}
\end{equation}
The mean impact velocity is independent of fragment dimensions.  Inserting
from Equations (46), (34), (31),~(33), (26), (47), (48), (19), and (15) into
Equation~(45), one finds that the collisional mean free time that is
consistent with the observational limitations, \mbox{$(\tau_{\rm coll})_{\rm
obs} = (\dot{N}_{\rm coll})_{\rm obs}^{-1}$}, equals{\vspace{-0.1cm}}
\begin{equation}
(\tau_{\rm coll})_{\rm obs} = \frac{\sqrt{5}}{30\eta \Phi}\:\!(G\rho X_{\rm
 frg})^{-\frac{1}{2}} D_{\textstyle \star}^{\frac{1}{4}} D_0^{-\frac{11}{4}}
 \! D_{\rm frg}^{\frac{7}{2}},
\end{equation}
where \mbox{$\eta \approx 1$}, \mbox{$G = 6.647 \!\times \!\!\!\:
10^7\!$\,cm$^3$\,g$^{-1}$\,yr$^{-2}$} is the{\nopagebreak} gravitational
constant, and
\begin{equation}
\Phi = \frac{1}{\Gamma_5} \!\left(\!1 \!+\! \frac{5\Gamma_3^2}{9\Gamma_5} \!
 \right) \!-\! \left(\! 1 \!+\! \frac{5\Psi_3^2}{9\Psi_5} \!\right) \!\!\cdot\!
 \!\left(\! 1 \!-\! \sqrt{\frac{D_{\textstyle \star}}{D_{\rm lim}}} \right) \!.
\end{equation}
Equation (50) replaces (35) as an expression for the mean free time between
collisions from the temporal distribution of the flare-ups observed by Liller
(2001).  Solving this equation for the cluster's collisional diameter $D_{\rm
frg}$, I list its values and the collisional parameters in Table 3 on the
{\vspace{-0.05cm}}assumptions that \mbox{$(\tau_{\rm coll})_{\rm obs} = 0.31$
yr},~\mbox{$\eta = 1$}, and \mbox{$D_{\rm lim} = 1.2$ km}.  In addition to $\langle \sigma \rangle$, $\langle V_{\rm rel}^2\rangle{\mbox{\raisebox{-0.4ex}{$^{\!^{1/2}
\!}$}}}$, and the true mean free time between collisions (i.e., both detected as
the flare-ups and undetected), $\tau_{\rm coll}$, I also list four key parameters
of the cluster of fragments from Table~1, as well as its average optical depth,
$\Theta$, defined as
\begin{equation}
\Theta = -\ln \!\left(\!1 \!-\! \frac{4X_{\rm frg}}{\pi D_{\rm frg}^2} \!
 \right) \!,
\end{equation}
and an average distance between the centers of neighboring fragments, $s_{\rm
frg}$, expressed by
\begin{equation}
s_{\rm frg} = \!\left( \! \frac{\pi \sqrt{2}}{6} \right)^{\!\!\frac{1}{3}}\!\!\!
 \cdot\!\left(\! \frac{D_{\textstyle \star}}{D_0} \!\right)^{\!\!\frac{5}{6}}
 \!\! D_{\rm frg} = 0.9047 \!\left(\!\frac{D_{\textstyle \star}}{D_0} \!
 \right)^{\!\! \frac{5}{6}} \!\! D_{\rm frg}.
\end{equation}
\begin{table}[b]
\vspace{-3.6cm}
\hspace{4.22cm}
\centerline{
\scalebox{1}{
\includegraphics{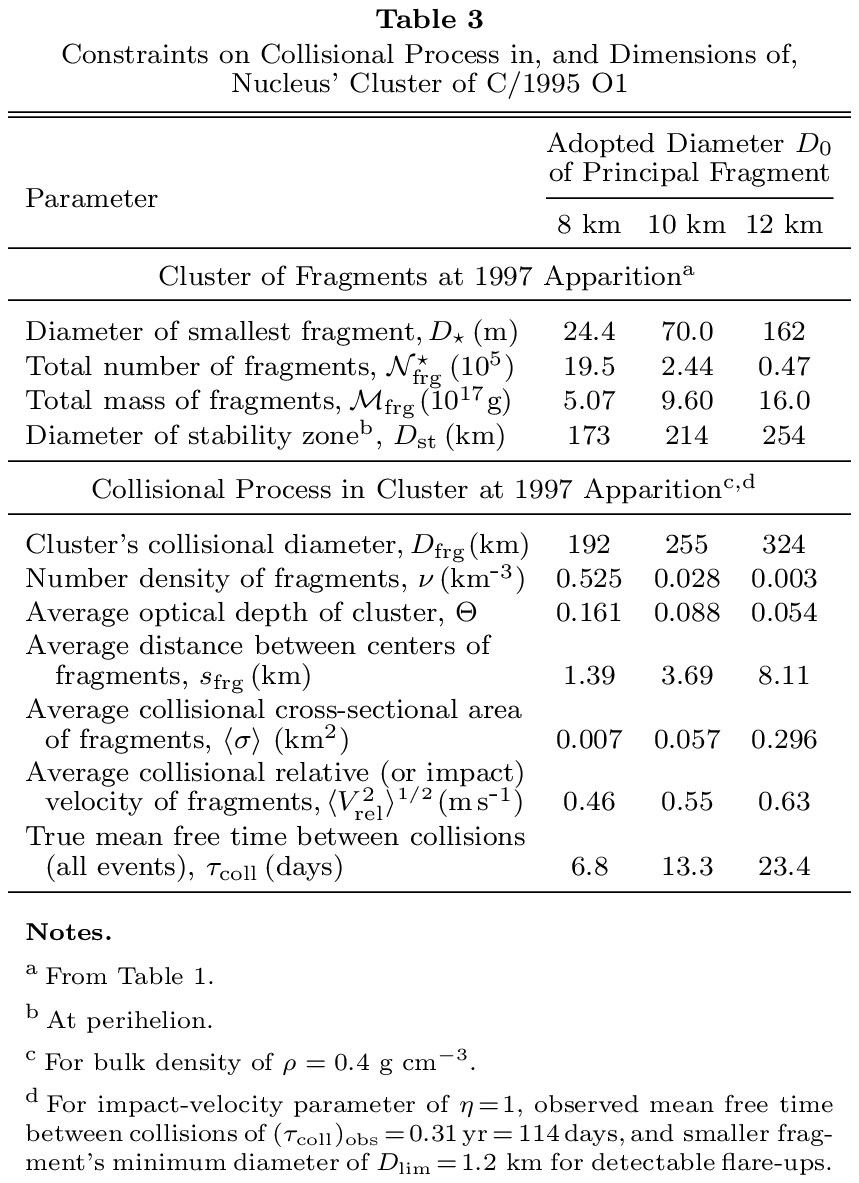}}} %  From: t3_HB2.tex
\vspace{-13.83cm}
\end{table}

Table 3 shows that for any of the three potential principal fragment's diameters
considered, the cluster's outer regions are exposed to the Sun's significant
perturbations at (and near) perihelion, as the collisional diameter then exceeds
the stability diameter.  It is expected that many fragments, especially at larger
distances from the center of mass, entered markedly different trajectories after
perihelion.  This development should clearly increase the fragments' orbital
diversity, thus the parameter $\eta$, and thereby give rise to a higher
collisional rate than before perihelion, which is consistent with Liller's prime
conclusion --- the absence of major preperihelion flare-ups.  Table 3 also suggests
that given the validity of the size distribution of fragments and the detection
limit, Liller observed, on the average, every fifth impact involving fragments
larger than $D_{\textstyle \star}$ in diameter if the principal fragment was
12~km across, but only every seventeenth impact if it was 8~km across.  

There are two additional more subtle effects mentioned by Liller (1997, 2001)
that could likewise be explained by the proposed hypothesis of a cluster-like
nucleus.  At heliocentric distances $r$ larger than $\sim$2.5~AU Liller fits
the quiescent phase of the light curve before and after perihelion by the same
power law, $r^{-n}$, where \mbox{$n = 2.55$}.  However, cursory inspection of
the light curve reveals that the post-perihelion data marginally deviate from
this slope, suggesting that the inner coma was fading at a slightly, but
perceptibly, lower rate than it was brightening before perihelion.  Yet, the
intrinsic brightness was nearly 0.2~mag higher before perihelion.  This behavior
is qualitatively consistent with two properties implied by the proposed
hypothesis:\ (i)~the comet continued to lose massive fragments from its
nucleus' cluster in the long run, hence it was brighter preperihelion; but
(ii)~the collisional rate was higher after perihelion (owing to the Sun's
major perturbations of fragments' trajectories around perihelion), hence some
of the new fragments lingered in the inner coma over longer periods of time
after perihelion and the inner coma's brightness was fading somewhat less
steeply.  Indeed, the post-perihelion normalized brightness was lower near
2.5~AU but caught up with the preperihelion brightness by $\sim$7~AU.

The other subtle peculiarity is Liller's (1997) reference to an apparent
quasi-periodic variability in the preperihelion light curve, with an average
period of \mbox{20$\,\pm\,$4}~days and a very small amplitude.  It is noted
from Table~3 that the {\it true\/} mean free time between {\it all\/} collisions
was as short as $\sim$7~days when one adopts a post-perihelion flare-up
triggering collisional rate of 3.2 per year.  However, I note that solutions
consistent with a $\sim$20~day periodicity require that the diameter $D_0$ of
the principal fragment not exceed about 11~km under any circumstances.  There
are no solutions fitting this periodicity for the larger dimensions.  

\begin{table}[b]
\vspace{-3.6cm}
\hspace{4.22cm}
\centerline{
\scalebox{1}{
\includegraphics{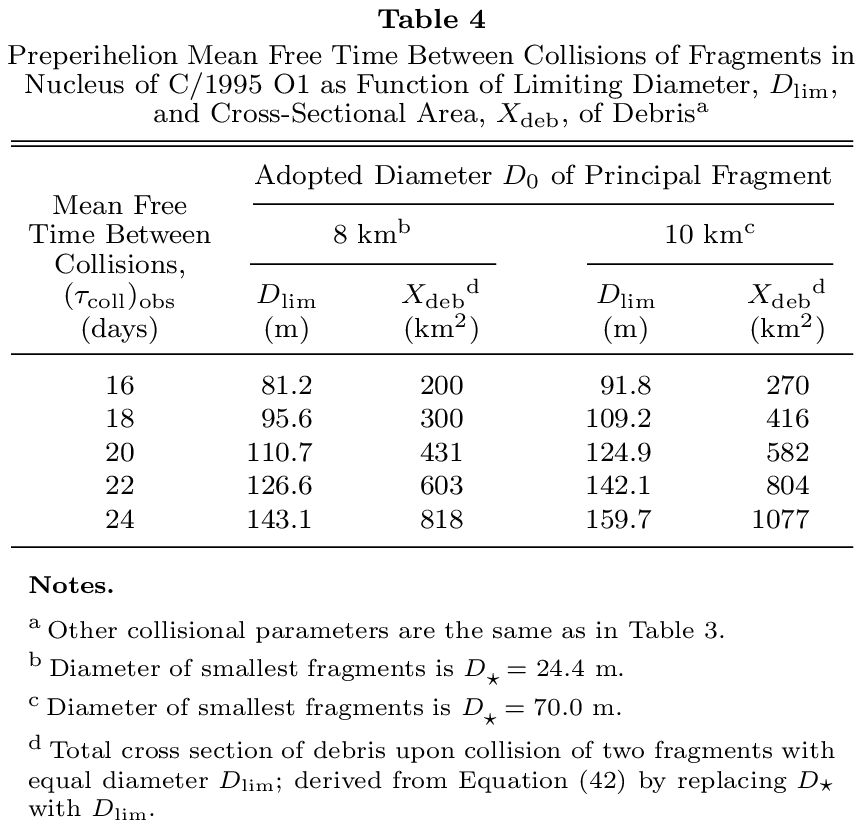}}} %  From t4_HB2.tex
\vspace{-17.43cm}
\end{table}

If the collisional rate was lower before perihelion, the 20-day period might
fit with \mbox{$D_{\rm lim} \!\sim\! D_{\textstyle \star}$}.  The dependence of
a mean free time between collisions of fragments on a cross-sectional area
of collisional debris and a limiting diameter $D_{\rm lim}$ is exhibited in
Table~4 for the principal fragment's adopted diameters of 8~km and 10~km.
The cross-sectional area of the debris is on the order of hundreds of square
kilometers only, too low to detect with Liller's instrumentation, and an
amplitude of the~\mbox{20-day} variations in the light curve, triggered
by the periodic presence of collisional debris in the inner coma, is estimated
to be merely on the order of thousandths of a magnitude.  The statistically
extracted amplitude of \mbox{0.1--0.2 mag}, apparent from Figure~5 of
Liller (1997), might be a product of the ensuing modest variations in the
carbon-monoxide production and in the associated ejection of microscopic dust,
as explained above.

\section{Resulting Constraints on Dimensions of\\Principal Fragment}
Up to Table 3, I executed all computations on three~different assumptions
regarding the diameter of the principal fragment --- 8~km, 10~km, and 12~km.
The tests~carried out in Sections~\mbox{2--9} and further evidence allow now
one to narrow down the width of the appropriate size span.  In the following
I separately discuss the various criteria in terms of the preferences for
particular segments of the 4-km wide range of the fragment's diameter.

\subsection{Nongravitational Acceleration}
The conservation-of-momentum criterion is of fundamental significance, because
it is on the strength of this evidence that the hypothesis of the comet's
nucleus in the form of a compact cluster of massive fragments has been
contemplated as the {\it only\/} credible scenario.  The estimated disparity
between the derived orbit-integrated nongravitational acceleration and the
one expected on the assumption that the nucleus was a single body of~the same
cross-sectional area amounts to more than two orders of magnitude in terms of
the momentum, or a little less than a factor of 10 in terms of the nucleus'
size.  The estimated uncertainty by a factor of \mbox{3--4} in the efficiency
of the momentum applied by outgassing is equivalent to an uncertainty by a
factor of only $\sim$1.5 in the linear dimension.  The principal fragment's
diameter of 12~km refers to a momentum transfer so efficient that it is barely
at a limit of feasibility; the lower end of the 4-km range for the principal
fragment's diameter (Table 3) should accordingly be assigned a much higher
probability.

\subsection{High-Resolution Imaging}
The high resolution imaging, especially with the HST, provides a strong argument
in favor of a very tight cluster.  Because the nucleus' image peaks up rather
sharply, its overall dimension can extend at most over just two pixels of the
HST's WFPC-2 sensor, each of which corresponds to \mbox{90--100}~km across on
four of the six analyzed preperihelion images between 1995 October 23 and 1996
October 17 (Sekanina 1999a), thus limiting the cluster's diameter to about
200~km at the extreme.  It is therefore {\it only\/} the lower end of the
principal fragment's range of dimensions that passes this test (Table~3).

\subsection{Stellar Occultation}
Fern\'andez et al.\ (1999) published the results of their campaign to
observe an occultation by the comet's nucleus of the star PPM\,200723 on
1996 October 5.  The event's light curve, obtained by apparently the only
team that reported a positive detection, was {\it V shaped\/}, without any
step-like variations or a trough implied by an occulting monolithic body
between the ingress and egress of the star.  Indeed, the best models offered by
Fern\'andez et al.\ included those with an occultation chord of zero length,
even though the degree of dimming indicated the star's complete disappearance
behind the comet at the time of maximum drop in the count rate.  Accordingly,
the authors could only conclude that the nucleus was less than 60~km in
diameter, an estimate that is too low and inconsistent with Szab\'o et al.'s
(2012) result.  For the nucleus in the form of a compact cluster of massive
fragments one expects a light curve that should be fairly smooth and
essentially trough-free.  The smaller the principal fragment's size, the
more consistent the cluster scenario is with the apparent absence of a
trough.  In spite of the uncertainties involved, the event appears to have
lasted \mbox{40$\,\pm\,$5 s}, suggesting the cluster's collisional diameter
of \mbox{204$\,\pm\,$26 km}, implying the principal fragment
\mbox{8.4$\,\pm\,$0.8 km} across.

\subsection{Sun's Perturbations of Cluster Fragments' Orbits}
Because the size of the zone of stability determined in Section~4 is approximate,
all that can be stated about the perturbation effects of the Sun near the 1997
perihelion is that they should have made the orbits of fragments near the outer
boundary of the cluster essentially chaotic.  As seen from Table~3,~the extent
of this instability increases with the size of the principal fragment more
steeply than linearly.  As a corollary,~the collisional rate among the fragments
in the cluster should have gotten augmented after perihelion regardless of the
principal fragment's size, as implied by Liller's (2001) results.

\subsection{Properties of Original Nucleus}
Listed in Table 1, the fundamental parameters of the original nucleus that is
presumed to have begun to fragment at the time of close encounter with Jupiter in
the year $-$2251 (or 2252 BCE) suggest, to a degree, that the principal fragment's
diameter of 8~km is more likely than 12~km.  Statistically, an original nucleus
22~km in diameter has a higher probability of occurrence than a nucleus 33~km
in diameter.  Similarly, the fragments would have stayed tighter together (and
the cluster would have had better gravitational stability) at a lower rotation
velocity (i.e., a smaller nucleus).  The lower end of the size range is also
more plausible because the central gravitational pressure is than more in line
with the compressive strength, estimated for 67P at \mbox{1--3}~kPa by Basilevsky
et al.\ (2016) and its upper limit at only 1.5~kPa by Groussin et al.\ (2015).
The latter team pointed out that diagenesis may then be initiated in the interior
of the comet's nucleus.  The probability of this process to have commenced in
C/1995~O1 increases with the square of the size of the original nucleus.  On the
other hand, the tensile strength needed for C/1995~O1 to begin to fragment at
the Jovian encounter just matched the lower end of the range reported for 67P,
thus making a larger original nucleus slightly more likely but by no means
indispensable.

\subsection{Distance Between Fragments in the Cluster}
It is noted from Table 3 that $s_{\rm frg}$, an average distance between the
centers of fragments, is in each of the three cases shorter than the diameter
of the principal fragment.  In fact, \mbox{$s_{\rm frg} \!\leq\! D_{\mbox{\tiny
\boldmath $\otimes$}} \leq D_0$} and the number of fragments whose diameters
are equal to or greater than $D_{\mbox{\tiny \boldmath $\otimes$}}$ is
\begin{equation}
{\cal N}_{\rm frg}(D_{\mbox{\tiny \boldmath $\otimes$}}) < \!\left( \!
 \frac{D_0}{s_{\rm frg}} \! \right)^{\!\!\frac{5}{2}}\!\!.
\end{equation}
It turns out that nearly 80 most massive fragments comply with this condition
when the principal fragment is 8~km across, 12 fragments when it is 10~km across,
but only the principal fragment and the second most massive fragment when 12~km
across.  This exercise provides a yet another argument for a high chance, if not
inevitability, of frequent collisions among fragments in the cluster.

In practice, this means that near the cluster's center of mass, where the
principal fragment and perhaps some other major fragments should reside, the
number density of fragments is much lower than the average, probably not more
than $\sim$10$^{-3}$ per km$^3$, given that the volumes of the 8~km, 10~km,
and 12~km fragments are, respectively, 268~km$^3$, 524~km$^3$, and 905~km$^3$.
Collisions of the principal fragment with other fairly large fragments are
likely to account for a fraction of the post-perihelion collisional rate
derived from Liller's (2001) observations.

\subsection{Impact Velocity and Collisional Rate}
Table 3 shows that an average impact velocity depends only moderately on
the principal fragment's dimensions and is close to 0.5~m~s$^{-1}$, high
enough to assure continuing fracture (rather than bouncing) of the initial
tidally-generated fragments of the original nucleus.  This expectation is
based on the assumption that, on the average, the impact velocities crudely
equal in magnitude the velocities of fragments about the center of mass of
the cluster in nearly circular orbits.  This is a plausible assumption, if
the fragments' motions are essentially random.  One deals here with a
self-feeding mechanism:\ the more often the collisions occur, the more
random the orbits become; the stochastic nature is also aided by the Sun's
perturbations, especially near perihelion.

Two further post-perihelion flare-ups were reported in gaps of Liller's (2001)
observing run, one in late August 1997 (McCarthy et al.\ 2007), four months before
Liller's first flare-up; the other in mid-October 1999 (Pearce 1999; Griffin \&
Bos 1999), six months after Liller's last flare-up.  If included, they would
increase $(\tau_{\rm coll})_{\rm obs}$ in Equation~(50) from 0.31~yr to 0.36~yr,
which would change the cluster's collisional diameter by $\sim$4\%, rather an
insignificant effect.

\subsection{A Verdict}
In summary, the lower end of the range of~the principal fragment's size --- a
diameter of \mbox{8--9}~km and a mass of \mbox{1.1--1.5$\,\times$10$^{17}$\,g}
--- comes out from this discussion as the preferred one by far, because of
the arguments presented primarily in Sections 10.1--10.2,
but~also~in~10.3,~and,~in part, in 10.5.  The cluster of fragments, some
\mbox{210$\,\pm\,$20 km} in~diameter,\,and~the~\mbox{pre-encounter} nucleus are
described by the data that can be interpolated from Tables~1~and~3.\hspace{0.1cm}

A complete summary of the cluster's adopted parameters for the principal fragment's
representative diameter of 8.5~km is tabulated in Section~12.  I may point out that
from their nongravitational model for comet motions and independently determined
nongravitational parameters (with the radial and transverse components only),
Sosa \& Fern\'andez (2011) derived for the nucleus of{\vspace{-0.05cm}}
C/1995~O1 a diamater of 9.6~km and a mass of \mbox{1.9$\,\times$10$^{17}$\,g},
remarkably close to the present results for the principal fragment.

For a cluster-like nucleus, many observed properties of C/1995 O1 (such as the
rotation vector, albedo, activity variations, complex dust-coma morphology and
its evolution, striation pattern in the tail, gravitationally-bound satellite,
etc.) will require a profound re-interpretation.  While a complete overhaul
of the large body of models that address these issues is outside the scope of
this paper, I pay attention to the problem of a satellite, a topic that turns
out to be particularly closely related to the proposed paradigm of a
compact cluster.

It is the orbital instability of fragments in the outer reaches of the cluster
--- which will from now on be referred to as a {\it primary (or main) nucleus'
cluster\/} or just a {\it primary\/} --- that lends legitimacy to such a link.
This instability virtually warrants that, from time to time, a fragment or a
subcluster of fragments escapes from the primary, thus contributing to a
population of boulder-sized debris scattered over an expanding volume of
hundreds or thousands of kilometers across around the primary.  A vast
majority of individual fragments are too feeble to detect even with the HST,
but subclusters of fragments should over a limited period of time show up as
faint companions.  This likely scenario invites a suggestion to conduct a
computer search for such companions in the HST images.
%
% In general, the separation of a satellite --- which itself might consist of a
% less massive compact cluster of fragments --- from the rest of the shattered mass
% of the original nucleus could be a product of a two-stage fragmentation process.
% Part of the original nucleus'  mass may have detached, because of the gradually
% increasing tidal forces, shortly before the closest approach to Jupiter,
% acquiring a slightly different orbital momentum, and settling into a less
% strongly bound orbit than the other ``parent'' fragments that were products
% of a subsequent event (or a sequence of events), nearer the time of closest
% approach.  This would be a less dramatic version of the same kind of process
% that caused the fragments of comet D/1993~F2 (Shoemaker-Levy) to line
% up as a string of beads along the orbit (e.g., Sekanina et al.\ 1998).

\section{On the Occurrence of Companion Nuclei}
My presentation of evidence on a major satellite orbiting the primary nucleus
of C/1995~O1 (Sekanina~1999b) has been a subject to controversy ever since,
in part~because the reported detection --- in five preperihelion images (in
1996 May--October) taken with the HST~Wide-Field Planetary Camera 2 (WFPC-2;
imaging scale~of 0$^{\prime\prime\!\!}$.0455 per pixel) through~an F675W
filter ---~was~made digitally; no companion was readily apparent in the
images when inspected visually.  The applied computer~pro\-cedure, based on
an iterative least-squares differential-correction technique, was described
in detail elsewhere (Sekanina 1995; an upgraded version in Sekanina 1999a).
The identification of this object as a satellite gravitationally bound to
the main nucleus was based on the assumption that the primary was at
least \mbox{3.4$\,\times$10$^{19}$\,g} in mass (i.e., 55~km across
at a bulk density of 0.4~g~cm$^{-3}$) and that therefore the radius of a
gravitationally stable zone around the nucleus was at least 370~km at
perihelion [Equation~(9) gives $\sim$350~km with \mbox{$h_0 = 0.1$}].  The
satellite, with its average projected distance of $\sim$180~km from the
primary, was safely inside the stability zone unless it was in each of the
five images located near the line of sight, a statistically unlikely scenario.

With a cluster of fragments replacing the solid nucleus, the situation is
rather different.  The stability zone being at perihelion less than 100~km
in radius (Table~1), the companion would have been located outside the
zone at heliocentric distances of up to at least $\sim$2~AU, i.e., over a period
of$\:\!\!${\gapeq}$\!\!$200~days around perihelion.  (On the other hand, the
companion should have remained inside the Hill sphere at all times.)  Dynamically,
it is possible but unlikely that these conditions would have sufficed for the
satellite to escape along an unbound orbit over a period of several
months.  It does not appear it happened, if the five satellite images refer
to the same object.  However, if the companion's existence dated back to the
close encounter with Jupiter, the 1997 perihelion was already a second instance
of severe solar perturbations.\footnote{This statement does not apply to the
published evidence on~the satellite based on the preperihelion observations
(Sekanina 1999b); however, as explained in Section~11.4, at least one major
companion was likewise detected in each of three post-perihelion HST images.}

\subsection{Doubts on the Existence of the Satellite}
In the past, dynamical issues were not at the focus~of a controversy on whether
the satellite (or companion){\nopagebreak} did in fact exist.  The doubts were
expressed because of the conditions under which the satellite's signature
was extracted from the HST images, in the presence of large amounts of dust
and its uneven spatial distribution in the inner coma of C/1995~O1.

Weaver \& Lamy (1999) questioned the detection on the grounds that the excess
signal attributed to the satellite is ``due to inadequate modeling of the
complex coma morphology and/or temporal variability.'' They also warned that
the HST's CCD arrays ``are imperfect detectors whose noise does not always
obey the laws of counting statistics,'' a problem that of course is model
independent.  In a follow-up paper, Weaver et al.\ (1999) reported that they
found no companions in the post-perihelion HST images taken in 1997--1998 with
a Space Telescope Imaging Spectrograph (STIS), but admitted that their detection
threshold was rather limited; I return to this topic in Section 11.4.  In his
review of the topics related to the size and activity of C/1995~O1, Fern\'andez
(2002) responded to the report of the satellite more suavely, pointing out that
the detection ``remains controversial because of the difficulty in understanding
the inner coma's brightness distribution.''{\vspace{-0.15cm}}

\subsection{Evidence Supporting a Companion's Existence}
A series of images of the comet was obtained on 1996 September 30 with a newly
commisioned adaptive optics system PUEO (also referred to as Bonnette, AOB) on the
{\vspace{-0.03cm}}Canada-France-Hawaii 3.6-m f/8 telescope on Mauna Kea (Rigaut et
al.\ 1998).\footnote{For the comet's images and their description, see {\tt
http://www.
cfht.hawaii.edu/$\!\!\:$Instruments/$\!\!\:$Imaging/$\!$AOB/best\_pictures.html.}}
When deconvolved, the images showed a ``knot of material'' 0$^{\prime\prime\!\!}$.15
north of the nucleus, at a position close to that of the reported satellite in an
HST image taken a week earlier (Sekanina 1999b).

Marchis et al.\ (1999) used another adaptive optics system, ADONIS, on the ESO
3.6-m telescope at La Silla, Chile, to take the comet's images on 1996 November
6 and 1997 January 15; in their deconvolved frames,~the central peak is clearly
resolved into two maxima of~uneven brightness, the fainter separated from the
brighter by 0$^{\prime\prime\!\!}$.23 at a position angle of
102$^\circ\pm\,$4$^\circ$ in November and by 0$^{\prime\prime\!\!}$.36 at
78$^\circ\pm\,$5$^\circ$ in January.  Marchis et al.\ considered three
interpretations for the secondary peak, including a companion nucleus, and
concluded that {\it based on their observations alone they remained undecided
as to whether the feature on either night was a near-nucleus footprint of a
jet or a secondary nucleus, but when combining their results with the findings
by others, the scenario involving} ``{\it a double nucleus \ldots seems to be
the most likely}.''~These authors also noted that in both cases the compact
feature projected close enough to the nucleus to qualify as~a gravitationally
bound object and that its position angles did not coincide with the directions
of the jets observed at 0$^\circ$--30$^\circ$, 75$^\circ$--95$^\circ$,
115$^\circ$--135$^\circ$, and 240$^\circ$--260$^\circ$ in November, and at
0$^\circ$--40$^\circ$ and 90$^\circ$--130$^\circ$ in January.{\vspace{-0.15cm}}

\subsection{Arguments and Counter-Arguments Based on Modeling Dust-Coma
 Morphology}
In a study of the dust-coma morphology of C/1995~O1 (Sekanina 1998), I pointed
out that --- given the comet's well determined spin vector --- a system of about
eight {\it evenly separated\/} halos in the southeastern quadrant of the coma,
prominently apparent in the comet's images from 1997 late February and early
March, could not (unlike the halos in the southwestern quadrant in the same
images) be modeled as dust ejecta from any source on the nucleus (not even on
the antisunward side) and that one has to admit that the observed morphology
is a product of dust ejecta from {\it two independent objects\/}\footnote{More
accurately, from {\it at least\/} two independent objects.} of different axial
orientation, thus providing further support for the existence of a companion.
This same conclusion was independently reached by Vasundhara \& Chakraborty
(1999) in their morphological study of C/1995~O1, and the argument was also
raised by Marchis et al.\ (1999).

On the other hand, Samarasinha (2000) argued that the discussed system of dust
halos in the southeastern quadrant could in fact be successfully modeled as a
product of an extended emission region (about 40$^\circ$ wide) on the surface of
the main nucleus.  Unfortunately, in an effort to demonstrate his idea, he applied
an approach that ignored effects of solar radiation pressure, an impermissible
omission.  As it turns out, an integrated contribution from the radiation pressure
became comparable in magnitude to the contribution from the ejection velocity ---
the variable that Samarasinha did account for in his model --- not later than in
the course of the third rotation (of the eight involved), but possibly earlier,
depending on the projected ejection velocity and the radiation pressure
acceleration of the submicron-sized grains that made up the features' outer
boundaries.  An even distribution of the consecutive halos in Samarasinha's
(2000) model is an artifact of his neglect of radiation pressure; its
incorporation into the model would compress the halos into a bright, extended
blob in the southeastern quadrant of the nucleus.  No such feature is apparent
in the comet's pertinent images, suggesting that the proposed active region of
enormous extent did not exist.

In general, the morphology of a dust-coma feature that is produced by an extended
source is modeled by using a collection of densely distributed point sources
(Sekanina 1987); this technique would certainly have been applied to C/1995\,O1
should it have been of any help. 

\subsection{Companions in HST's STIS Images}
In Section 11.1 I remarked on three post-perihelion images of C/1995~O1 taken in
1997--1998 with a then~newly installed STIS instrument on board the HST as well
as on Weaver et al.'s (1999) report of their non-detection~of any companions.  In
this paper I present for the first~time the results of my subsequent search for
companions in these images, using the technique that was previously applied to the
HST's preperihelion WFPC-2 images (Sekanina 1999a, 1999b).   The advantages of
STIS over \mbox{WFPC-2} are a higher quantum efficiency and a lower readout~noise
of its CCD array and much broader imaging passbands, thus reaching objects
$\sim$1.5 mag fainter.  On the other hand, STIS is less well photometrically
defined than WFPC-2, with its point spread function expected to degrade
approximately 30\% near the boundary of the field of view (Baum
1996).{\vspace{-0.2cm}}

\begin{table*}[t]
\vspace{-4.12cm}
\hspace{-0.53cm}
\centerline{
\scalebox{1}{
\includegraphics{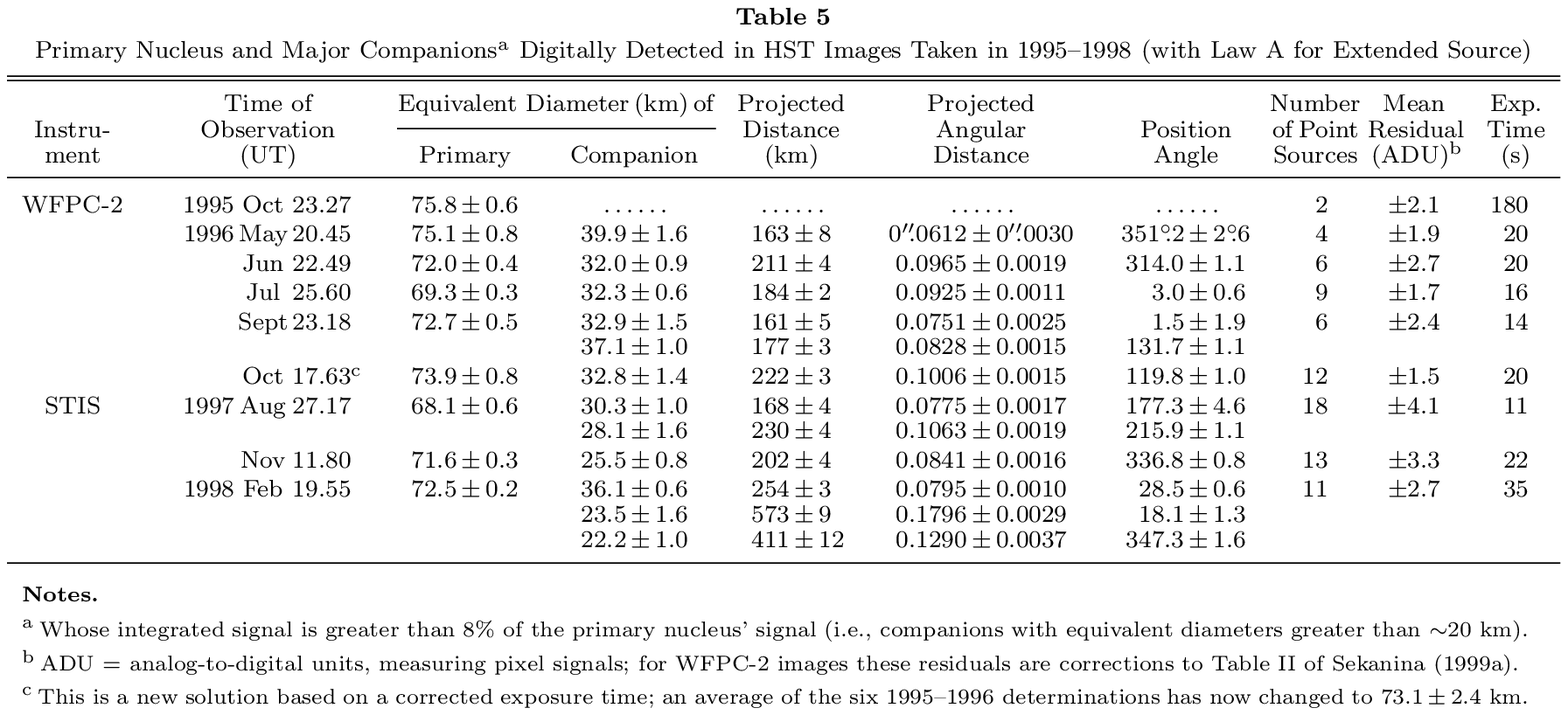}}} %  From: t5_HB2.tex
\vspace{-16.52cm}
\end{table*}

\subsubsection{Method of CCD signature extraction}
The observed surface brightness distribution was available as an array of pixel
signals measured in CCD analog-to-digital intensity units (ADU px$^{-2}$), with
{\vspace{-0.045cm}}each pixel 0$^{\prime\prime\!\!}$.0508 on a side.  A background
noise of 3~ADU px$^{-2}$ was subtracted.  The net pixel signals were assumed to
consist of a convolved sum of three contributions:\ (i)~from one or more extended
sources (to model the coma's complex morphology); (ii)~from the primary nucleus
(dominant point source); and (iii)~from additional point sources, some of which
could represent genuine companions (nuclear fragments or their clusters), while
others are fictitious spots of light of instrumental or unknown origin.

Employing its surface brightness map for STIS images, the point spread function
(PSF) was approximated by a quasi-Gaussian law with a symmetrical surface
brightness distribution $b_{\rm psf}(X,Y)$, which at a point \{$X,Y$\}, whose
distance from the PSF's peak at \{$X_{\textstyle \star}, Y_{\textstyle \star}$\}
equaled \mbox{$\Delta X = X \!\!-\!\! X_{\textstyle \star}$} and \mbox{$\Delta
Y = Y \!\!-\!\! Y_{\textstyle \star}$}, was expressed as
\begin{equation}
b_{\rm psf}(X,Y) = b_{\textstyle \star} \exp \! \left[ - \! \left( \! \frac{\Delta
 X^2 \!+ \Delta Y^2}{2 \sigma_{\rm psf}^2} \! \right)^{\!\!\nu_{\rm psf}} \right],
\end{equation}
where $\sigma_{\rm psf}$ is the PSF's dispersion parameter, $\nu_{\rm psf}$ is a
dimensionless constant, and \mbox{$b_{\textstyle \star} = b_{\rm psf}(X_{\textstyle
\star}, Y_{\textstyle \star})$} is the peak surface brightness.  The integrated
brightness $I_{\textstyle \star}$ of the point source is then
\begin{equation}
I_{\textstyle \star} = 2 \pi b_{\textstyle \star} \sigma_{\rm psf}^2
 \nu_{\rm psf}^{-1} \Gamma\:\!\!\!\left(\! \nu_{\rm psf}^{-1} \!\right),
\end{equation}
where $\Gamma(z)$ denotes the Gamma function of argument~$z$.  For the
long-pass (LP) filter the parametric values are \mbox{$\sigma_{\rm psf} =
0.1461$\,px} and \mbox{$\nu_{\rm psf} = 0.3034$}, so that $I_{\textstyle
\star}$ in ADU is
\begin{equation}
I_{\textstyle \star} = 1.181 b_{\textstyle \star},
\end{equation}
where $b_{\textstyle \star}$ is in ADU px$^{-2}$.  Each point source is
fully~described by three constants:\ $X_{\textstyle \star}$, $Y_{\textstyle
\star}$, and $I_{\textstyle \star}$.

A surface brightness distribution in extended sources, $b_{\rm ext}(X,Y)$, was
approximated (after its convolution with the PSF) by an ellipsoidal power law
(referred to as law A in Sekanina 1999a, 1999b), which allowed for a deviation
of the peak's location in the ellipsoid's center, described by \{$X_{\rm ext},
Y_{\rm ext}$\}, from the origin of the coordinate system, as well as for anisotropy
and an arbitrary orientation, the latter defined by an angle $\theta_{\rm ext}$
in the direction of the most gentle rate of signal decline from the peak:
\begin{equation}
b_{\rm ext}(X,Y) = \frac{b_0}{1 \!+\! \left[ \!\left( \! {\displaystyle
 \frac{\Delta X}{\sigma_x}} \!\right)^{\!\!2} \!\!+\! \left( \! {\displaystyle
 \frac{\Delta Y}{\sigma_y}} \!\right)^{\!\!2} \right]^{\!\frac{1}{2}\nu_{\rm
 ext}}} \, ,
\end{equation}
where
\begin{equation}
\left(\!\!
\begin{array}{c}
\Delta X \\ \Delta Y
\end{array}
\!\!\right) \!=\! \left(\!\!
\begin{array}{cc}
\cos \theta_{\rm ext} & \sin \theta_{\rm ext} \\
-\sin \theta_{\rm ext} & \cos \theta_{\rm ext}
\end{array}
\!\!\right) \!\cdot\! \left(\!\!
\begin{array}{c}
X\!\!-\!\!X_{\rm ext} \\ Y\!\!-\!\!Y_{\rm ext}
\end{array}
\!\!\right) \!,
\end{equation}
$\sigma_x$ and $\sigma_y$ are the maximum and minimum dispersions of the
surface brightness distribution along, respectively, the $X$ and $Y$ axes, and
$\nu_{\rm ext}$ is the exponent of the power law.  Each extended source is
fully described by seven~independent constants:\ $X_{\rm ext}$, $Y_{\rm ext}$,
$\sigma_x$, $\sigma_y$, $b_0$, $\nu_{\rm ext}$,~and~$\theta_{\rm ext}$.

To summarize, in a search for $N_{\rm pt}$ point sources and $N_{\rm ext}$
extended sources, the deconvolving procedure's optimization least-squares
differential-correction technique was required to iteratively solve for
\mbox{$3 N_{\rm pt} \!+\! 7 N_{\rm ext}$} parameters; typically, signals
in 157 pixels were fitted.{\vspace{-0cm}}

\begin{figure}
\vspace{-1.65cm}
\hspace{0.26cm}
\centerline{
\scalebox{0.525}{
\includegraphics{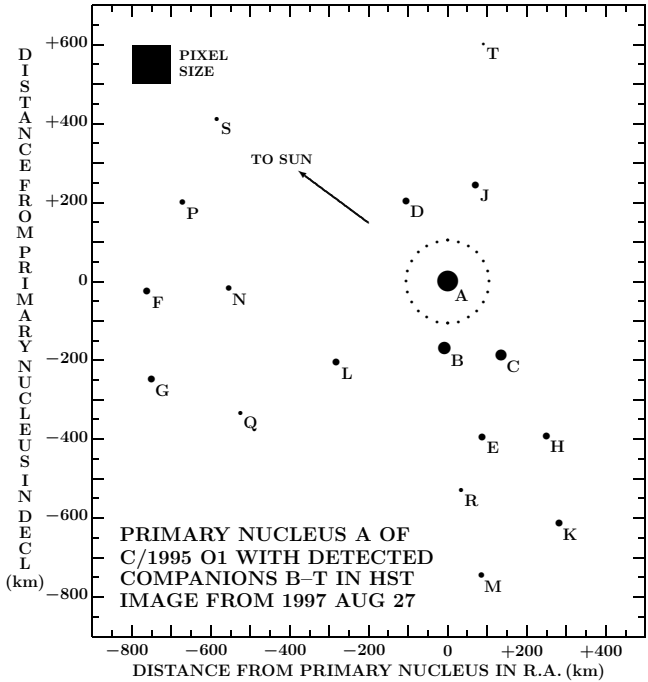}}} %  From: f2_HB2.tex
\vspace{-5.1cm}
\caption{Spatial distribution of detected companions B--T relative to the
primary nucleus A of comet C/1995~O1 in projection onto the plane of the
sky on 1997 August 27.  The size of the circles is to scale on the
assumption of infinite opacity, i.e., for an equivalent diameter listed in
Table~6.  The dotted circle shows the dimensions of the compact cluster of
fragments that represents the primary nucleus A, as adopted in Section~10.8.
The diameters of the companion nuclei should be adjusted proportionately,
if they too are judged to consist of compact clusters of fragments.
Unlike dust jets, most companions project relative to the primary nucleus
in directions that are far from the direction to the Sun.{\vspace{0.4cm}}}
\end{figure}

\subsubsection{Results}
The extracted dimensions of the primary nucleus and some results of a search
for companions in the STIS images are presented in Table~5 together with
the~partially revised results of a previous analysis of the preperihelion
WFPC-2 images\footnote{The image from 1996 October~17 was reanalyzed, because
only after the original papers (Sekanina 1999a, 1999b) were accepted~for
publication it became known that the initially announced exposure time was
incorrect (see Table~I in Sekanina 1999a).  In addition, recent inspection of
the computer runs revealed that the mean~residuals in Table~II of Sekanina
(1999a) were inadvertently multiplied by a factor of~10, an error that has now
been corrected in Table~5.} (Sekanina 1999a, 1999b).  The tabulated numbers and
experience with the fitting{\nopagebreak} procedure offer these conclusions:
(1)~The fitted dimensions{\nopagebreak} of the primary nucleus are remarkably
consistent over the orbital arc of 28~months, with an {\it equivalent diameter\/}
(measuring the observed cross-sectional area) averaging \mbox{73.1$\,\pm\,$2.4 km},
within 0.4$\sigma$ of the result derived by Szab\'o et al.\ (2012); (2)~introduction
of a second extended source in the solutions consistently failed~to improve the
fit to the distribution of dust in the coma, implying --- together with a low rms
residual of about $\pm$2--4~ADU --- that the employed distribution~function given by
Equation~(58) provided an adequate approximation; (3)~a large number of companion
nuclei was detected with both instruments, the larger post-perihelion numbers being
owing in part to the STIS instrument's higher sensitivity; (4)~as shown in an
example~in Figure~2, most companion nuclei~were~not,~unlike dust jets, located in
directions close to that of the Sun\footnote{Contained in the 90$^\circ$ sector
centered on the projected sunward direction are only five of the 17~companions in
the 1997 August 27 image; only five of the 12~companions in the 1997 November 11
image; and --- astonishingly --- none of the 10~companions in the 1998 February
19 image.} and were not concentrated densely along particular lines, thus making
their interpretation as phenomena that were closely related to dust jets quite
unlikely; (5)~as further documented by \mbox{Tables~6--8}, the brighter companions
had an extremely high signal-to-noise ratio close to or exceeding 10, and only
for a few of the tabulated objects could their existence be readily questioned,
in particular, the companions R, S, and T in Table~6, N$^\prime$ in Table~7, and
K$^{\prime\prime}$ and L$^{\prime\prime}$ in Table~8.

\begin{table}[t]
\vspace{-4.08cm}
\hspace{4.25cm}
\centerline{
\scalebox{1.0}{
\includegraphics{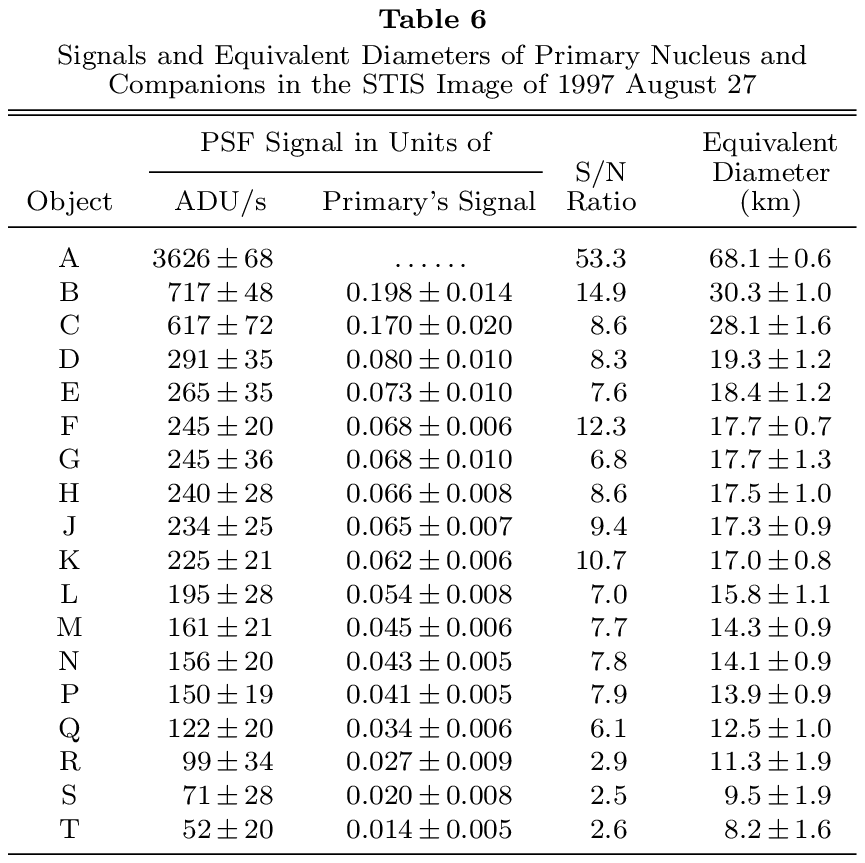}}} %  From: t6_HB2.tex
\vspace{-16.6cm}
\end{table}
\begin{table}[b]
\vspace{-3.85cm}
\hspace{4.25cm}
\centerline{
\scalebox{1.0}{
\includegraphics{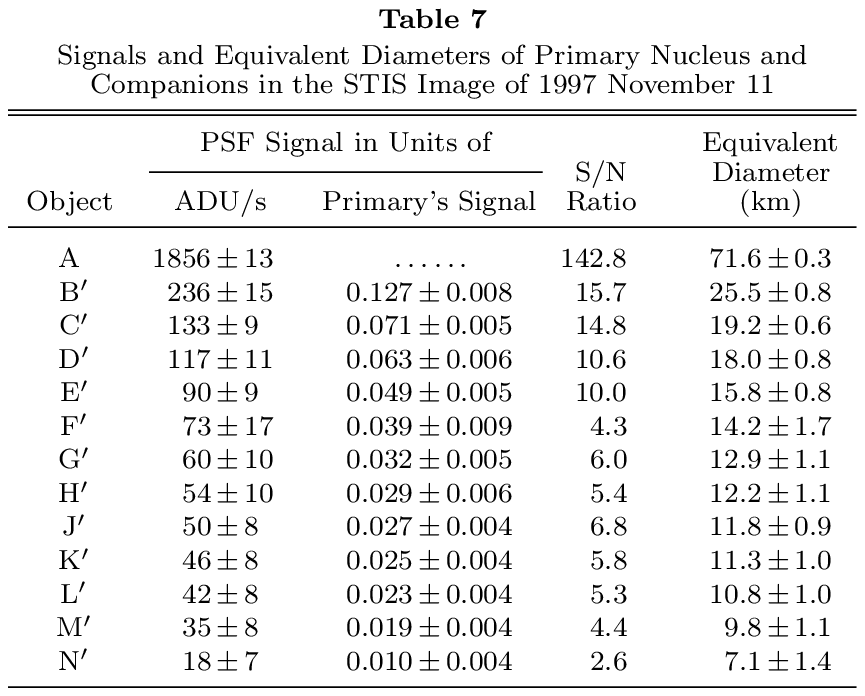}}} %  From: t7_HB2.tex
\vspace{-18.8cm}
\end{table}

The primary nucleus (i.e., the compact cluster of fragments proposed
to make it up) is marked A in all three STIS images; \mbox{B--T} are the
companion nuclei detected in the image taken on 1997 August 27 (Table~6);
\mbox{B$^\prime$--N$^\prime$} the companions detected in the image of 1997
November 11 (Table~7); and \mbox{B$^{\prime\prime}$--L$^{\prime\prime}$}
the companions in the image of 1998 February 19.  As pointed out, the
 existence of the objects R, S, T, N$^\prime$, K$^{\prime\prime}$, and
L$^{\prime\prime}$ is questionable; the existence of G$^{\prime\prime}$,
H$^{\prime\prime}$, and J$^{\prime\prime}$ is somewhat uncertain.

It may be significant that the bright companion in the image of 1996
October~17 (Table~5), taken only 20 days before the ESO observation that
we referred to in Section~11.2, is located at a position angle that differs
from that in the ESO image by less than 4$\sigma$ and at a comparable
angular distance from the primary nucleus.  The positional difference
may be due in part to the companion's motion in the course of the 20~days,
in part to effects introduced by the heavy processing of the ESO image.

\begin{table}[t]
\vspace{-4.1cm}
\hspace{4.25cm}
\centerline{
\scalebox{1}{
\includegraphics{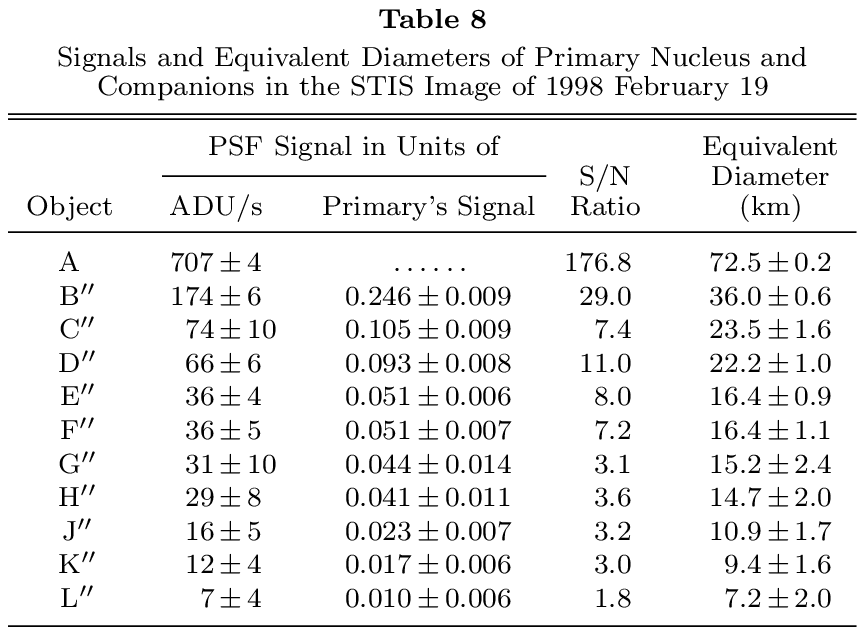}}} %  From: t8_HB2.tex
\vspace{-18.85cm}
\end{table}

A list of all detected companion nuclei is in the order of increasing projected
distance from the primary nucleus presented in Table~9.  Since the radius of the
stability zone defined by{\vspace{-0.04cm}} Equation~(9) (with \mbox{$h_0 = 0.1$}
and the primary nucleus' mass of 6$\,\times$10$^{17}$g; cf.\ Section 12)
amounted to 250~km for the image of 1997 August 27, to 340~km for the image of
1997 November 11, and to 450~km for the image of 1998 February 19, only 3--6
innermost companions were likely to have been, on any of the three dates, located
inside the stability zone of the primary.  On the other hand, all companions were
located deep inside the Hill sphere of the primary nucleus --- whose radii at the
three times were between 1700~km and 3100~km --- unless the distances were in
each image strongly foreshortened.  This result suggests that the primary nucleus
was still likely to exert much influence over the motions of many if not all of
the detected companions at the times of observation.

\begin{figure}[b]
\vspace{-1.9cm}
\hspace{0.57cm}
\centerline{
\scalebox{0.52}{
\includegraphics{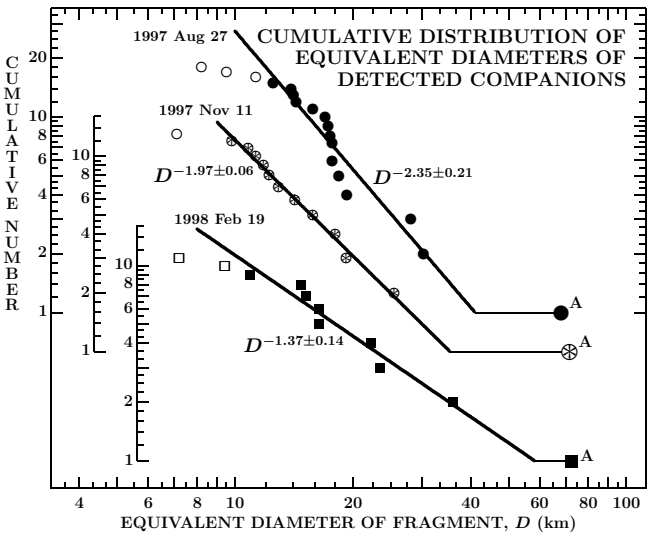}}} %  From: f3_HB2.tex
\vspace{-6.1cm}
\caption{Cumulative distribution of equivalent diameters of the detected
companions in the STIS images.  The primary nucleus~(A) and the companions 
of questionable existence (signal-to-noise ratio of $\leq$3 in Tables~6--8),
shown by open symbols on the left, deviate from the distribution and were
not employed in the fitting.  The slope of the distribution drops rapidly
with time.{\vspace{-0.1cm}}}
\end{figure}
\begin{table}[t]
\vspace{-4.1cm}
\hspace{4.25cm}
\centerline{
\scalebox{1}{
\includegraphics{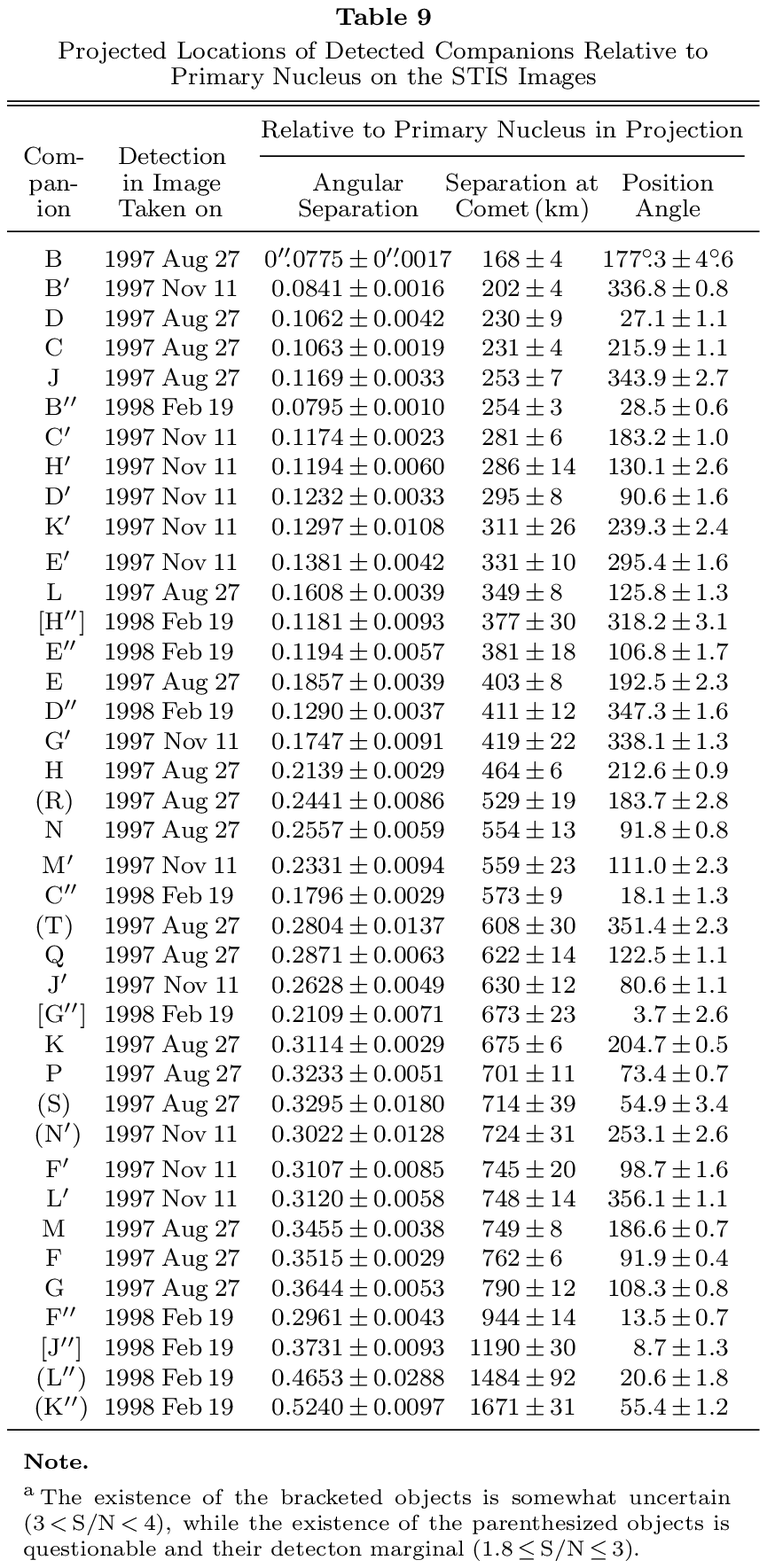}}} %  From: t9_HB2.tex
\vspace{-7.50cm}
\end{table}

Figure 3 displays the cumulative distribution of the detected companions'
equivalent diameters, which measure, as in the instance of the primary nucleus,
their observed cross-sectional areas.  A striking property of the distribution
is a very rapid rate of slope flattening with time, from $D^{-2.35 \pm 0.21}$
in August 1997 to $D^{-1.97 \pm 0.06}$ in November 1997, to $D^{-1.37 \pm 0.14}$
in February 1998.  As the expected slope of a steady-state distribution law for
fragments of a common parent is $D^{-2.5}$, described by Equation~(12), the trend
in the slope in Figure~3 points to a peculiar behavior, to be addressed below.
Furthermore, the equivalent diameters of most companions are so large that they
cannot be single fragments; instead, they appear to consist of subclusters of
fragments.

The data points in Figure 3 that significantly deviate from the fitted
power laws include the companions~whose existence is questionable; this is
understandable, because their signals barely exceeded the background signal
of the extended source.  Not only was their detection marginal, but there may
have existed additional companions of similar eqivalent diameters that failed
to be detected at all, an argument that is in agreement with the positions
of the doubtful companions consistently below the fitted laws in the figure.
Alternatively, of course, there may not be any companions comparable in size
or smaller than the questionable ones and the population of companions may
terminate right there.  Also deviating substantially from the fitted laws is
the primary nucleus, especially in the first two of the STIS images; fairly
large offsets are not unusual at the lower end of the cumulative distributions
of statistical sets and are not necessarily worrisome.

As for the rapidly dropping slope of the cumulative distribution of the
companions in Figure~3, I test an assumption that it is caused by temporal
variations in the objects' cross-sectional area (measured by their equivalent
diameter).  The issue is important because if the companions are compact
subclusters of fragments, their total cross sections may either increase
with time as a result of progressive fragmentation because of collisions at very
low velocities --- or decrease also as a result of the fragmentation that entails
escape of much of the involved mass from the gravity field of the subcluster
once the debris acquired velocities that exceeded the escape limit.  Which of
the two processes dominates is determined by the systematic variations in the
distribution's slope, linked to the cross-sectional variations with time.

I now examine under what set of circumstances could the steady-state cumulative
distribution of companions, given by Equations~(13), change dramatically
its slope to fit the distributions in Figure~3.  Let at time $t_0$, when
the process affecting the distribution of companions was set off, an
equivalent diameter of the primary nucleus be \mbox{$D_0^{\textstyle \ast} =
D_0^{\textstyle \ast}(t_0)$}, while a companion's equivalent diameter be
\mbox{$D^{\textstyle \ast} \!= D^{\textstyle \ast}(t_0) = D_1^{\textstyle
\ast}(t_0), D_2^{\textstyle \ast}(t_0)$, \ldots \,[$D^{\textstyle \ast}(t_0)
\!<\! D_0^{\textstyle \ast}(t_0)$]}.  Calling \mbox{$x = x(t_0) = D^{\textstyle
\ast}(t_0)/D_0^{\textstyle \ast}(t_0)$}, the steady-state cumulative distribution,
${\cal N}_{\rm nuc}$, of the companion nuclei at time $t_0$ is, following
Equation~(13),
\begin{equation}
{\cal N}_{\rm nuc}(t_0) = x^{-\kappa_0},
\end{equation}
where \mbox{$x \!\leq\! 1$}, \mbox{$\kappa_0 \!=\! \frac{5}{2}$}, \mbox{${\cal
N}_{\rm nuc} \!=\! 1$} for \mbox{$x \!=\! 1$} (when \mbox{$D^{\textstyle \ast}
\!=\! D_0^{\textstyle \ast}$}), and \mbox{${\cal N}_{\rm nuc} \!>\! 1$} for
\mbox{$x \!<\! 1$}.

At a time \mbox{$t > t_0$}, a different relationship~applies,~as~is demonstrated
by Figure~3.  In particular, one~now~has \mbox{$D_0^{\textstyle \ast} \!=\!
D_0^{\textstyle \ast}(t)$},\,\mbox{$D^{\textstyle \ast} \!=\! D_1^{\textstyle
\ast}(t),\,D_2^{\textstyle \ast}(t)$},\,\ldots{\vspace{-0.02cm}}[with~\mbox{$
D_0^{\textstyle \ast}(t) \!\neq\! D_0^{\textstyle \ast}(t_0)$},
\mbox{$D_i^{\textstyle \ast}(t) \neq D_i^{\textstyle \ast}(t_0), \,i = 1, 2$,
\ldots]}, {\vspace{-0.02cm}}\mbox{$x = x(t) = D^{\textstyle
\ast}(t)/D_0^{\textstyle \ast}(t)$}, and, generally, \mbox{$x(t) \neq x(t_0)$}
with the exception of \mbox{$x = 1$}.  The cumulative distribution at time $t$
is described by
\begin{equation}
{\cal N}_{\rm nuc}(t) = x^{-\kappa},
\end{equation}
where \mbox{$\kappa \neq \kappa_0$} and constraints similar to those in
Equation~(60) apply to ${\cal N}_{\rm nuc}$.

The issue now is to modify Equation~(60) in a way such that it describes the
cumulative distribution of equivalent diameters observed at time $t$ and
simultaneously reproduces the distribution in Equation~(61).  I search for
a solution by adding a function $y(x)$, to be determined, to the variable $x$
from Equation~(61), so that Equation~(60) appears at time $t$ as follows:
\begin{equation}
{\cal N}_{\rm nuc}(t) = (x \!+\! y)^{-\kappa_0}.
\end{equation}
The function $y$ is subject to a boundary condition \mbox{$y = 0$} at
\mbox{$x = 1$} in order that \mbox{${\cal N}_{\rm nuc} = 1$}.  The log-log
derivative~of the expression (62) becomes
\begin{equation}
\frac{d \log {\cal N}_{\rm nuc}}{d \log x} = \frac{x}{{\cal N}_{\rm nuc}}
 \frac{d {\cal N}_{\rm nuc}}{dx} = -\kappa_0 \, \frac{x}{x \!+\! y}
 \left( \! 1 \!+\! \frac{dy}{dx} \right) \!,
\end{equation}
while from Equation~(61) one gets immediately
\begin{equation}
\frac{d \log {\cal N}_{\rm nuc}}{d \log x} = -\kappa.
\end{equation}
Comparing the right-hand sides of the expressions (63) and (64), one obtains
a linear differential equation of the first order,
\begin{equation}
\frac{dy}{dx} - \chi \frac{y}{x} = \chi \!-\! 1,
\end{equation}
where \mbox{$\chi \!=\! \kappa/\kappa_0 \!<\! 1$} because \mbox{$\kappa \!<\!
\kappa_0$} from Figure~3.  The general solution to Equation~(65) is
\begin{equation}
y(x) = c_0 x^\chi - x,
\end{equation} 
where $c_0$ is a constant; from the boundary condition~for~$y$ in Equation~(62)
one finds \mbox{$c_0 = 1$}, so that
\begin{equation}
y(x) = x \, (x^{\chi-1} \!-\! 1).
\end{equation}
Inserting from the solution (67) for $y$ into Equation~(62), one indeed obtains
at once Equation~(61).  Since \mbox{$\chi \!<\! 1$}, $y$ is positive for any
\mbox{$x \!<\! 1$} and, in conformity with Figure~3, the expression (67)
implies that \mbox{$x(t) \!<\! x(t_0)$} for any \mbox{${\cal N}_{\rm nuc}
\!>\! 1$}; the process of accelerating escape of fragments from the companion
clusters, entailing a progressively~increasing loss of their cross-sectional
area, dominates.

The observed slopes from Figure 3 can be fitted as a function of heliocentric
distance $r$, expressed in units of peri\-helion distance $q$, by an exponential
law of the type
\begin{equation}
\kappa(r) = \kappa_0 \exp \!\left\{ C_1 \!\left[ 1 \!-\! \left( \frac{r}{q}
 \right)^{\!\!C_2} \right] \! \right\},
\end{equation}
where $C_1$ and $C_2$ are constants.  A fairly broad range~of the pairs $C_1$,
$C_2$ fits the slopes of the three observed distributions from Figure~3 about
equally well, with the~re\-sulting residuals much smaller than the errors
involved.{\hspace{0.3cm}}

\begin{table}[b]
\vspace{-3.5cm}
\hspace{4.03cm}
\centerline{
\scalebox{0.975}{
\includegraphics{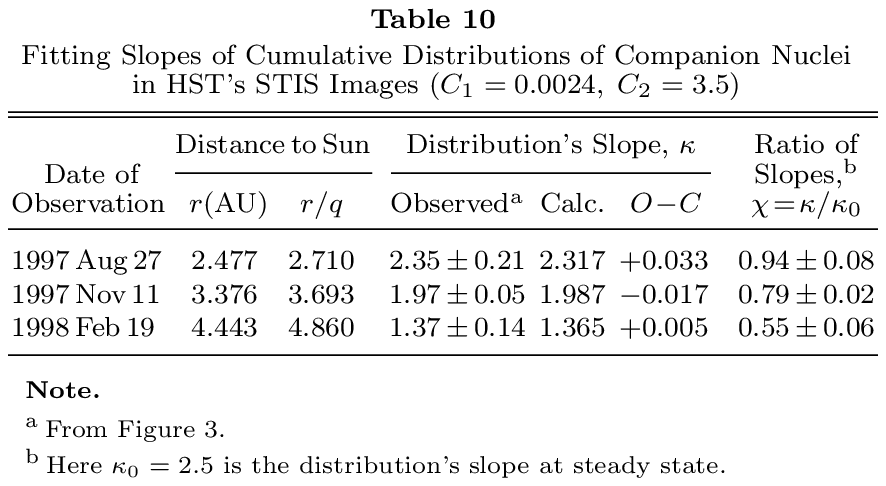}}} %  From: t10_HB2.tex
\vspace{-20.52cm}
\end{table}

Table 10 presents the calculated values of the slopes~$\kappa$ and their
residuals from one particular fit that employs{\nopagebreak} \mbox{$C_1 =
+0.0024$} and \mbox{$C_2 = +3.5$}.  Note that for \mbox{$r \!=\! q$},
Equation~(68) always satisfies a condition \mbox{$\kappa = \kappa_0 = 2.5$}
regardless of the choice for $C_1$ and $C_2$.  It thus appears that the gradual
loss of the cross-sectional area of the detected companions may have been
triggered off by the Sun's significant near-perihelion perturbations of the
motions of fragments in each companion's cluster, with some of them presumably
lost to space at an accelerated rate after perihelion.  The time of birth of the
companions is unknown but the results suggest that the distribution of these
subclusters of fragments still conformed to steady state at the time of
perihelion passage.

If correct, this argument implies that the {\it preperihelion\/} distribution
of equivalent diameters of companions was essentially in steady state.  To
test this inference, I examined the distribution for the image of 1996
October 17, in which 11 companions were detected,  a greater number than
in any other preperihelion image.  I found that, once again, with the
exception of the primary nucleus and a few companions of questionable
existence at the other end of the distribution, the equivalent diameters of
the remaining 8 companions fitted a power law with a slope of \mbox{$\kappa
= 2.67 \pm 0.22$}, consistent within errors with the steady-state slope
of 2.5.

A general picture of the distribution of equivalent diameters of the companions
that emerges from these considerations is a possible cyclic variation in the
slope:\ the assemblage of companions approaches perihelion with a steady-state
distribution, but departs it with a distribution that is increasingly flatter.
To retain this cycle from one revolution about the Sun to the next, the
process of slope flattening is required to terminate at some time after
perihelion and steady state to be gradually restored.  It is probable that
far from the Sun it is the collisions of fragments in the cluster of each
companion (as well as the primary nucleus) that in the absence of the Sun's
perturbations gradually re-establish steady state within the primary's cluster
and, by extension, in the distribution of the companions' equivalent diameters
as well.  If so, the rapidly diminishing tilt of the distribution in an early
post-perihelion span of time is merely a short-lived solar-perturbation
effect.

Alternatively, the rapidly flattening slope of the distribution of the
companions' equivalent diameters could represent a lasting effect that
wiped out the steady-state distribution once and for all.  If this
interpretation is correct, it would imply the birth of the detected
companions to date back to a time (or times) {\it after\/} the previous
perihelion in the year of $-$2250, as the steady-state distribution
should otherwise have been done away with shortly after that time.

In either scenario, the practical issue is the degree of contamination caused
by the companions in the primary nucleus' signal detected in the images taken
at very large heliocentric distances after perihelion that Szab\'o et
al.\ (2012) used in their investigation.  For example, within a projected
distance of 500~km of the primary nucleus, the companions contributed 71\%
of the primary nucleus' signal in the 1997 August image (Table~6), 40\%
in the 1997 November image (Table~7), and 39\% in the 1998 February image
(Table~8).  If this trend continued, the contribution became probably
close to negligible in an HST image of 2009 September 8, one of the images
employed by Szab\'o et al.  However, since the degree of contamination
depends on the instrumental constants, it may represent one of possible
explanations for a minor gap of 0.19 mag between the 2009 HST observation
and the 2011 October 23 VLT observations, which Szab\'o et al.\ attributed
to an albedo difference.

In closing, a correlation is noticed in this context that appears to
exist after perihelion between the decreasing slope of the distribution
curve of the companions' equivalent diameters and the presence of the
striking flare-ups on the comet's inner-coma light curve, observed by
Liller (2001) (Section~9).  Either phenomenon is proposed to be a
signature of fragment collisions, in both the primary nucleus' cluster
and the subclusters of the companions.

\subsection{Final Comments on the Problem of\\Companion Nuclei in C/1995~O1}
The first comment concerns the terminology.  In my earlier paper (Sekanina
1999b) I consistently referred to a {\it satellite\/} or {\it satellites\/},
whereas now I am dealing with a {\it companion\/} or {\it companions\/}.  As
remarked at the beginning of Section~11, this change of terminology is a corollary
of the new model for the comet's nucleus that implies a substantially lower mass
of the cluster that makes up the primary nucleus, by a factor of more than 100,
relative to the mass estimated for a solid nucleus of the same cross-sectional
area.  This difference clearly has an effect on the dimensions of the stability
zone, with the result that the range of distances for gravitationally bound
companions --- the satellites --- is now curtailed significantly.

The second comment is to underscore a point that appears to have never been
contemplated in the controversy of the detection of a companion (or companions)
in close proximity of the primary nucleus:\ the unequal degrees of fitness
and resolution offered by the various applied techniques toward achieving
a detection.  I argue that a two-dimensional modeling of the type that the
method employed here is based on is more robust and less prone to missing
inconspicuous objects in close proximity of a major object than is, for
example, the method of radial cuts used in Lamy's approach (e.g., Lamy et
al.\ 1996).

The superior qualities of the applied technique are apparent not only from
the high signal-to-noise ratios of the detected companions (as listed in
Tables~6--8), but also from the results of experimentation with fitting
additional extended sources to account for a complex distribution of the
signal over the investigated field of view.  Solutions with more than one
extended source were not successful in fitting the local signal peaks, unlike
the solutions with additional point sources.  Thus, the present results
cast doubts on the detected bumps in the digital maps as imprints of complex
morphological features of the ambient dust coma (such as jets or hoods) and,
instead, support the notion that they are signatures of point-like companion
nuclei immersed in the coma.

\section{Summary and Conclusions}
A consensus is that C/1995~O1 was one of the most spectacular comets of
the 20th century with an unusually large nucleus.  However, even the best
determination --- based on the far-infrared observations with the Herschel
Space Observatory when the comet was no longer active --- entailed
assumptions by employing a model that converted the observed flux, that is,
a measure of the {\it cross-sectional area\/}, of 4300 km$^2$, to what
should be called an equivalent diameter of \mbox{74\,$\pm$\,6 km} (Szab\'o
et al.\ 2012).

Accordingly, it is the cross-sectionial area --- a quantity more directly
measured than the diameter --- that describes the nucleus more faithfully.
It is then a matter of interpretation to decide what kind of nucleus the
measured quantity characterizes, whether a single spherical solid body, or
a binary object, or a cluster of solid spherical bodies of the same overall
cross sectional area, etc.  Subject to additional constraints, they all
satisfy the flux condition equally well.

It is noted that Szab\'o et al.'s result is in excellent agreement with a
mean equivalent diameter of 73.1\,$\pm$\,2.4~km (Table~5), derived from the
HST images taken on nine dates between 1995 October and 1998 February, when the
comet was always less than 6.4~AU, and as close as 2.7~AU, from the Sun.  This
correspondence suggests that the comet's dust coma was at these heliocentric
distances optically thin all the way to the surface of the nucleus.

A gigantic size of the nucleus is fundamentally at odds with the independent
detection of a fairly high outgassing-driven nongravitational acceleration that
the comet's orbital motion was subjected to{\vspace{-0.04cm}} (Paper~1).  The
acceleration,{\vspace{-0.04cm}} (0.707\,$\pm$\,0.039)$\times 10^{-8}$AU~day$^{-2}$
at a heliocentric distance of 1~AU and equal to (2.39\,$\pm$\,0.13)$\times 10^{-5}$
the Sun's gravitational acceleration, follows a modified Marsden-Sekanina-Yeomans
(1973) law with a scaling distance of \mbox{$r_0 \!=\!  15.36$}~AU.  When
integrated over the entire orbit about the Sun, the nongravitational effect is
found to be equivalent to a momentum change per unit mass of
2.46\,$\pm$\,0.14~m~s$^{-1}$.  Accounting for the momentum exerted by the
mass sublimated from the nucleus over the orbit, the conservation-of-momentum
law suggests that such a nucleus of a bulk density of 0.4~g~cm$^{-3}$ should
not exceed $\sim$10~km in diameter and its mass should be on the order of
10$^{17}$g, more than {\it two orders of magnitude\/} less than the mass of
the nucleus with the dimensions determined by Szab\'o et al.\ (2012).

I argue that this major conflict can only be avoided by postulating that the
nucleus of C/1995~O1 at its recent return to perihelion was made up of a {\it
compact cluster of massive fragments\/} of the original nucleus that broke up by
the action of tidal forces exerted by Jupiter during the comet's close encounter
with the planet in the 23rd century BCE (Paper~1).  Dominated by collisions, the
cluster is assumed to have a size distribution of fragments that reached steady
state, with their cumulative number varying inversely as a $\frac{5}{2}$th power
of fragment diameter.  The nongravitational acceleration detected in the comet's
orbital motion is in this scenario interpreted as referring to the principal,
most massive fragment, located near the cluster's center of mass and, as required
by the conservation-of-momentum law, up to 10~km in diameter.  The nongravitational
accelerations on other~outgassing fragments remain undetected, triggering
perturbations of their motions relative to the principal fragment.

Besides having a correct cross-sectional area, the cluster ought to appear as a
nearly point-like feature in the high-resolution images taken with the HST
instruments.  This requisite limits the models to a strongly compacted cluster not
exceeding $\sim$200~km in diameter, constraining its image's extent to no more
than two pixels across on the HST detectors at geocentric distances of $\sim$3~AU.
Independently, the steady-state size distribution of fragments restricts the total
mass of the cluster to less than five times the mass of the principal fragment.

Further critical properties of C/1995~O1 are the tensile strength of its
original, pre-encounter nucleus as well as a degree of gravitational stability
and collisional rate of the cluster-like nucleus.  The mass of the pre-encounter
nucleus is estimated at about 20 masses of the principal fragment and more
than 20~km in diameter, with most of the mass having been lost by the time the
comet was discovered in 1995.  Given the minimum encounter distance of less than
11~Jovian radii (Paper~1), a critical tensile strength along fissures could not
be higher than several Pa for the nucleus to fracture, while the central
gravitational pressure did not exceed $\sim$3000~Pa.

Based on existing studies of gravitational stability of globular clusters on
the one hand and binary asteroids on the other hand, I adopt for the radius
of a stability zone a conservative limit equaling $\sim \! \frac{1}{7}$th the
radius of the Hill sphere.  For a cluster of fragments of the considered mass,
the sphere of gravitational stability at perihelion of C/1995~O1 is slightly
smaller than the cluster's dimensions.  Accordingly, significant perturbations
of the cluster's outer reaches by the Sun are likely near perihelion, resulting
presumably in a higher collisional rate after perihelion, at least over limited
periods of time.

Possible evidence for this corollary of the Sun's near-perihelion perturbations
is Liller's (2001) list of recurring flare-ups in the comet's inner coma, five
of which were detected between early 1998 and mid-1999.  An additional flare-up
of similar nature was observed by~other astronomers in late 1999, when Liller's
monitoring was incomplete.  Interpreting the flare-ups as due to dust ejecta
from colliding kilometer-sized fragments, Liller's result provides their
collisional rate, allowing thus to correlate the cluster's dimensions with the
principal fragment's size and to select a narrow range of cluster models centered
on the most probable one, presented in Table~11.  The high chance of collisions
among fragments is illustrated by an average distance between their centers,
which is shorter than the diameters of the $\sim$50 most massive fragments.

The proposed cluster model for the nucleus of comet C/1995~O1 so dramatically
contrasts with the traditional single-body model that the published interpretations
of the comet's coma morphology and brightness variations have now become largely
invalidated and will have to be reinvestigated essentially from scratch, an effort
that is beyond the scope of this study.  However, the flare-ups in the
post-perihelion light curve of the inner coma (Liller 2001) are unlikely to be
products of sudden local activity on a single rotating nucleus because identical
areas of the surface would have been exposed to the Sun before perihelion,
yet no flare-ups were detected along the incoming branch of the orbit.

\begin{table}[t]
\vspace{-4.2cm}
\hspace{4.22cm}
\centerline{
\scalebox{1}{
\includegraphics{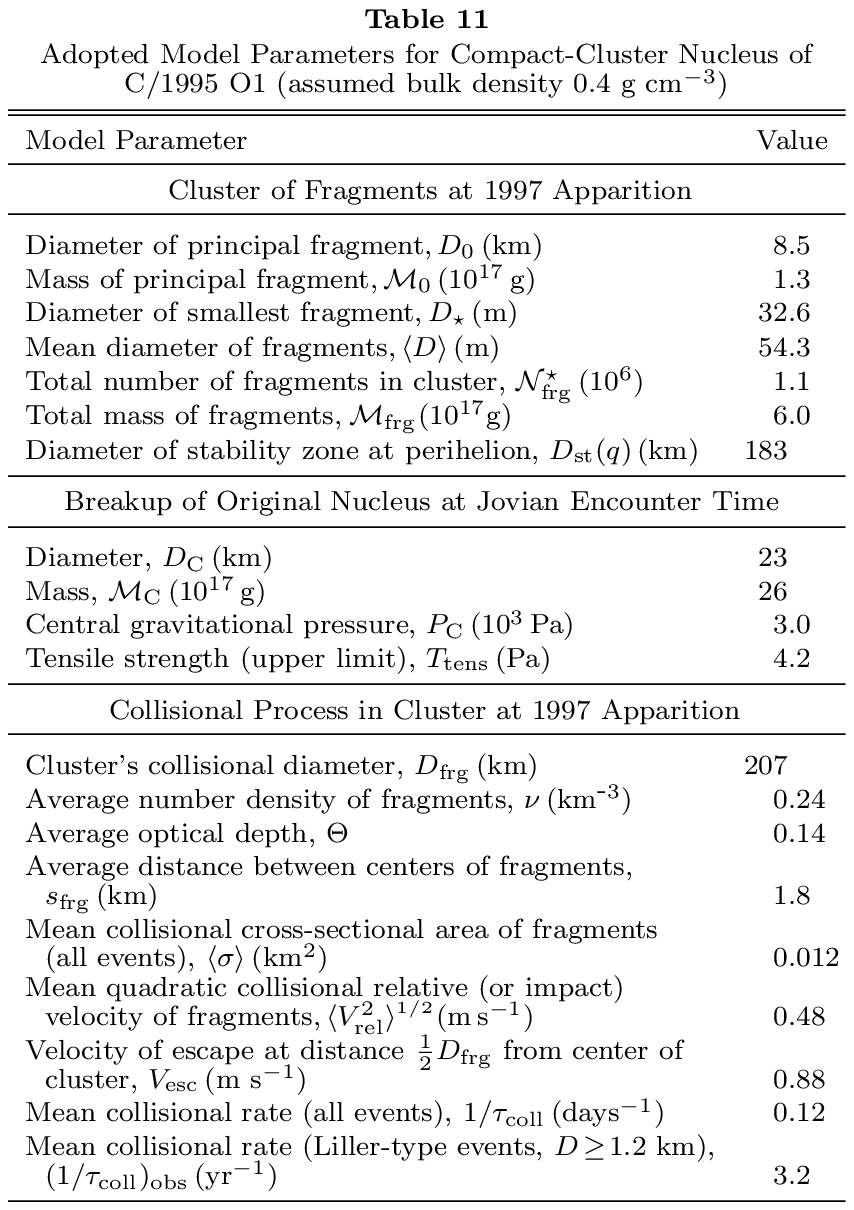}}} %  From: t11_HB2.tex
\vspace{-13cm}
\end{table}

Also beyond this investigation's scope is a highly desirable Monte Carlo-type
modeling of the fragments' motions in the nucleus cluster over long periods of
time, involving an n-body problem.  This task should be undertaken to assess
the degree of the cluster's gravitational stability along the preperihelion
leg of the orbit, the severity of the Sun's perturbations especially around
perihelion given that the size of the stability zone is then found to be
slightly smaller than the cluster's size, and the magnitude of their effects
on the motions of individual fragments inside or outside the cluster along
the post-perihelion leg of the orbit.

A final comment on the compact-cluster model of the nucleus of C/1995~O1
relates to a major imbalance between the masses of the original, pre-encounter
nucleus and the cluster structure at the 1997 apparition, by which time the
comet is estimated to have lost about three quarters of its initial mass,
most of it in direct orbits apparently soon after the encounter.  It is
expected that attrition was also likely to accompany the process
of perturbing the motions of fragments near the 1997 perihelion.  Given the
submeter-per-second velocities of fragments, this scenario invites a suggestion
that a fraction of the perturbed fragments or subclusters of fragments escaped
the main cluster's gravity shortly following the perihelion passage and that
such boulder-sized debris should be scattered near the comet and might show
up in the HST's post-perihelion images from 1997 August 27 through 1998
February 19.  Indeed, a fragment moving with a velocity of escape radially
away from the primary nucleus' cluster in free flight would be expected to
reach a distance of $\sim$1000~km from it in a matter of a few weeks; in
reality, the time needed to reach this distance would be longer by a factor
of a few.  And closer objects may represent stray fragments or subclusters
of fragments whose relative velocities were just below the velocity of escape.
Inspection of the immediate proximity of the primary nucleus for such stray
objects was therefore eminently desirable and a computer search harvested
more than 30 objects in the three post-perihelion HST images and at least
15 in the first of them alone.

The detected signals of these companions suggest that they in fact are
subclusters of fragments because, if single objects, most of them would be
larger than the principal fragment of the primary nucleus' cluster.  Their
signal-to-noise ratios are generally quite high, and the possibility that
they all are factitious products of the search algorithm is so remote that
it can safely be dismissed.  Also, there is no correlation between their
locations and the positions and directions of bright jet-like features in
the images.  As to the old controversy on the detection of a companion (or
satellite) in the HST images, it could very well be that it is the applied
computer technique's faculty, aptitude, and sensitivity to extracting the
object's signal that makes the difference, a point never raised in the past.

While the times of the post-perihelion HST images were judged to be separated
by gaps too wide to establish identities of the companions in different images,
the data set was considered appropriate for investigating the cumulative
distribution of their signals (or equivalent diameters) separately in each
frame.  The result was a surprisingly rapid systematic drop in the slope of
the distribution from a rate lower than, but fairly close to, the steady-state
rate in the 1997 August 27 image to a rate about 1.7 times less steep in the
1998 February 19 image.  The trend is explained as an effect of a gradual
dissipation of the subclusters that make up the companions, whereby the more
massive subclusters decay at a slower pace than the less massive ones.
Quantitative analysis implies that when extrapolated back in time, the
steady-state distribution would have been reached near perihelion, a
coincidence that suggests a possible relationship between the separation of
the subclusters from the primary and the Sun's peak perturbations.

I do realize that the developed compact-cluster~model, while self-consistent,
might be deemed by some~as~controversial and certainly in need of further
testing.~Similarly, some of the comet's properties might be hard to make readily
compatible with this model or they might require additional constraints or
introduce further conditions.  Whatever difficulties might lie ahead,\,however,\,the
major disparity between the outgassing-driven \mbox{nongravitational} acceleration
--- which for C/1995~O1 exceeds, or is comparable to, nongravitational
accelerations derived in~the past for fairly bright, but by no means
spectacular,~long-period comets\footnote{For example, for comet{\vspace{-0.04cm}}
C/1998~T1 the total nongravitational parameter amounted to 0.88\mbox{$\:\!$}$\times
10^{-8}$ AU day$^{-2}$ (Nakano {\vspace{-0.04cm}}2000); for C/2000 WM$_1$ to
0.52\mbox{$\:\!$}$\times 10^{-8}$ AU day$^{-2}$ (Nakano 2002);{\vspace{-0.04cm}}
and for the main component B of C/2001~A2 to 0.59\mbox{$\:\!$}$\times 10^{-8}$
AU day$^{-2}$ (Nakano 2001). (The errors are in the second or higher decimal.)}
--- and the single-body model, which for {\vspace{-0.04cm}}C/1995~O1 requires a
nucleus 74~km in diameter~and nearly 10$^{20}$\,g in mass, strikes one as so
utterly compelling that to me this argument alone rules the traditional,
single-body model out completely.  And I am aware of no other model that would
stand.\\[0.17cm]

I thank Franck Marchis for providing me with the digital maps of the three
post-perihelion HST images made with the STIS instrument in 1997--1998.
This research was carried out at the Jet Propulsion Laboratory, California
Institute of Technology, under contract with the National Aeronautics and
Space Administration.\\[-0.2cm]
\begin{center}
\raisebox{0.03cm}{\footnotesize REFERENCES}
\end{center}
\vspace*{-0.3cm}
\begin{description}
{\footnotesize

% \item[\hspace{-0.3cm}]
% Acree, W., \& Chickos, J. S. 2016, J. Phys. Chem. Ref. Data, 45,{\linebreak}
%  {\hspace*{-0.6cm}}033101
% \\[-0.57cm]
%
\item[\hspace{-0.3cm}]
Aggarwal, H. R., \& Oberbeck, V. R. 1974, ApJ, 191, 577
\\[-0.57cm]
\item[\hspace{-0.3cm}]
Basilevsky, A. T., Krasil'nikov, S. S., Shirayev, A. A., et al.\ 2016,{\linebreak}
 {\hspace*{-0.6cm}}Solar Sys. Res., 50, 225
\\[-0.57cm]
\item[\hspace{-0.3cm}]
Baum, S. 1996, STIS Instrument Handbook, Version 1.0.  (Balti-{\linebreak}
 {\hspace*{-0.6cm}}more:\ Space Telescope Science Institute)
\\[-0.57cm]
\item[\hspace{-0.3cm}]
Bessel, F. W. 1836, AN, 13, 185
\\[-0.57cm]
\item[\hspace{-0.3cm}]
Biver, N., Bockel\'ee-Morvan, D., Crovisier, J., et al.\ 2002, Earth{\linebreak}
 {\hspace*{-0.6cm}}Moon Plan., 90, 5
\\[-0.57cm]
%
% \item[\hspace{-0.3cm}]
% Bockel\'ee-Morvan,\,D., Crovisier,\,J., Mumma,\,M.\,J., \& Weaver,\,H.\,A.{\linebreak}
% {\hspace*{-0.6cm}}2004, in Comets II, ed.\,M.\,C.\,Festou,\,H.\,U.\,Keller,\,\&\,H.\,A.\,Weaver{\linebreak}
% {\hspace*{-0.6cm}}(Tucson, AZ: University of Arizona), 391
% \\[-0.57cm] % composition of comets
%
\item[\hspace{-0.3cm}]
Brownlee, D. E. 1985, Annu.\ Rev.\ Earth Plan.\ Sci., 13, 147
\\[-0.57cm]
%
% \item[\hspace{-0.3cm}]
% Cercignani, C. 1981, Progr.\ Astronaut.\ Aeronaut., 74, 305
% \\[-0.57cm]
\item[\hspace{-0.3cm}]
Chebotarev, G. A. 1964, Sov.\ Astron., 7, 618
\\[-0.57cm]
\item[\hspace{-0.3cm}]
Crovisier, J., Bockel\'ee-Morvan, D., Colom, P., et al.\ 2004, A\&A,{\linebreak}
 {\hspace*{-0.6cm}}418, 1141
\\[-0.57cm]
\item[\hspace{-0.3cm}]
Della Corte, V., Rotundi, A., Fulle, M., et al. 2015, A\&A, 583, A13
\\[-0.57cm]
\item[\hspace{-0.3cm}]
Dohnanyi, J. S. 1969, JGR, 74, 2531
\\[-0.57cm]
\item[\hspace{-0.3cm}]
Fern\'andez, Y. R. 2002, Earth Moon Plan., 89, 3
\\[-0.57cm]
\item[\hspace{-0.3cm}]
Fern\'andez, Y. R., Wellnitz, D. D., Buie, M. W., et al.\ 1999,~Icarus,{\linebreak}
 {\hspace*{-0.6cm}}140, 205
\\[-0.57cm]
\item[\hspace{-0.3cm}]
Fulle, M., Cremonese, G., \& B\"{o}hm, C.\ 1998, AJ, 116, 1470
\\[-0.57cm]
\item[\hspace{-0.3cm}]
Griffin, I. P., \& Bos, M. 1999, IAUC 7288
\\[-0.57cm]
\item[\hspace{-0.3cm}]
Groussin, O., Jorda, L., Auger, A.-T., et al.\ 2015, A\&A 583, A32
\\[-0.57cm]
\item[\hspace{-0.3cm}]
Gunkelmann, N., Ringl, C., \& Urbassek, H. M. 2016, A\&A, 589,{\linebreak}
 {\hspace*{-0.6cm}}A30
\\[-0.57cm]
%
% \item[\hspace{-0.3cm}]
% Hale, A., \& Bopp, T. 1995, IAU Circ., 6187
% \\[-0.57cm]
%
\item[\hspace{-0.3cm}]
Hamilton, D. P., \& Burns, J. A. 1991, Icarus, 92, 118
\\[-0.57cm]
\item[\hspace{-0.3cm}]
Hamilton, D. P., \& Burns, J. A. 1992, Icarus, 96, 43
\\[-0.57cm]
\item[\hspace{-0.3cm}]
Jorda, L., Lamy, P., Groussin, O., et al.\ 2000, in ISO Beyond Point{\linebreak}
 {\hspace*{-0.6cm}}Sources:\ Studies of Extended Infrared Emission,
 ESA-SP 455, ed.{\linebreak}
 {\hspace*{-0.6cm}}R.\,J.\,Laureijs,\,K.\,Leech, \&\,M.\,F.\,Kessler\,(Noordwijk,\,Netherlands:{\linebreak}
 {\hspace*{-0.6cm}}ESTEC), 61
\\[-0.57cm]
\item[\hspace{-0.3cm}]
Kennedy, G. F. 2014, MNRAS, 444, 3328
\\[-0.57cm]
\item[\hspace{-0.3cm}]
Kessler, D. J. 1981, Icarus, 48, 39
\\[-0.57cm]
\item[\hspace{-0.3cm}]
Kramer, E. A., Fern\'andez, Y. R., Lisse, C. M., et al.\ 2014, Icarus,{\linebreak}
 {\hspace*{-0.6cm}}236, 136
\\[-0.57cm]
\item[\hspace{-0.3cm}]
Lamy, P. L., Toth, I., Gr\"{u}n, E., et al.\ 1996, Icarus, 119, 370
\\[-0.57cm]
%
% \item[\hspace{-0.3cm}]
% Lamy, P. L., Jorda, L., Toth, I., et al.\ 1999, BAAS, 31, 1116
% \\[-0.57cm]
%
\item[\hspace{-0.3cm}]
Lamy, P. L., Toth, I., Fern\'andez, Y. R., \& Weaver, H. A. 2004, in{\linebreak}
 {\hspace*{-0.6cm}}Comets II,\,ed.\,M.\,C.\,Festou,\,H.\,U.\,Keller,\,\&\,H.\,A.\,Weaver\,(Tucson:{\linebreak}
 {\hspace*{-0.6cm}}University of Arizona Press), 223
\\[-0.57cm]
\item[\hspace{-0.3cm}]
Licandro, J., Bellot Rubio, L. R., Boehnhardt, H., et al.\ 1998,
 ApJ,{\linebreak}
 {\hspace*{-0.6cm}}501, L221
\\[-0.65cm]
\item[\hspace{-0.3cm}]
Liller, W. 1997, Plan.\ Space Sci., 45, 1505}
% \\[-0.57cm]
%
\end{description}
\pagebreak
\vspace*{0.3cm}
\begin{description}
{\footnotesize
\item[\hspace{-0.3cm}]
Liller, W. 2001, Int.\ Comet Q., 23, 93
\\[-0.57cm]
\item[\hspace{-0.3cm}]
Marchis,\,F.,\,Boehnhardt,\,H.,\,Hainaut,\,O.\,R.,\,\&\,Le\,Mignant,\,D.~1999,
 {\hspace*{-0.6cm}}A\&A, 349, 985
\\[-0.57cm]
\item[\hspace{-0.3cm}]
Marcus, J. N. 2007, Int.\ Comet Q., 29, 39
\\[-0.57cm]
\item[\hspace{-0.3cm}]
Mardling, R. A. 2008, in The Cambridge N-Body Lectures, Lecture{\linebreak}
 {\hspace*{-0.6cm}}Notes in Physics, ed.\  S.\ J.\ Aarseth, C.\ A.\ Tout, \&
 R.\ A.\ Mardling{\linebreak}
 {\hspace*{-0.6cm}}(Berlin:\ Springer), 760, 59
\\[-0.57cm]
\item[\hspace{-0.3cm}]
Marsden, B. G. 1968, AJ, 73, 367
\\[-0.57cm]
\item[\hspace{-0.3cm}]
Marsden, B. G. 1969, AJ, 74, 720
\\[-0.57cm]
%
% \item[\hspace{-0.3cm}]
% Marsden, B. G. 1970, AJ, 75, 75
% \\[-0.57cm]
%
% \item[\hspace{-0.3cm}]
% Marsden, B. G. 1999, Earth Moon Plan., 79, 3
% \\[-0.57cm]
%
% \item[\hspace{-0.3cm}]
% Marsden, B. G., \& Williams, G. V. 2008, Catalogue of Cometary
%  {\hspace*{-0.6cm}}Orbits 2008, p.\ 108 (17th ed.; Cambridge, MA:
%  Smithsonian {\hspace*{-0.6cm}}Astrophysical Observatory, 195pp)
% \\[-0.57cm]
%
\item[\hspace{-0.3cm}]
Marsden, B.\,G., Sekanina, Z., \& Yeomans, D.\,K.\ 1973, AJ,\,78,\,211
 \\[-0.57cm]
%
% \item[\hspace{-0.3cm}]
% Marsden, B. G., Sekanina, Z., \& Everhart, E. 1978, AJ, 83, 64
% \\[-0.57cm]
%
\item[\hspace{-0.3cm}]
McCarthy, D. W., Stolovy, S., Campins, H., et al.\ 2007, Icarus, 189,{\linebreak}
{\hspace*{-0.6cm}}184
\\[-0.57cm]
\item[\hspace{-0.3cm}]
Merouane, S., Zaprudin, B., Stenzel, O., et al.\ 2016, A\&A,~596,~A87
\\[-0.57cm]
\item[\hspace{-0.3cm}]
Nakano, S. 2000, Minor Plan.\ Circ.\ 40668
\\[-0.57cm]
\item[\hspace{-0.3cm}]
Nakano, S. 2001, Minor Plan.\ Circ.\ 44030
\\[-0.57cm]
\item[\hspace{-0.3cm}]
Nakano, S. 2002, Minor Plan.\ Circ.\ 46619
\\[-0.57cm]
\item[\hspace{-0.3cm}]
Pearce, A. R. 1999, IAUC 7288
\\[-0.57cm]
\item[\hspace{-0.3cm}]
Probstein, R. F. 1969, in Problems of Hydrodynamics~and~Con-{\linebreak}
 {\hspace*{-0.6cm}}tinuum Mechanics, ed.\ F. Bisshopp \& L. I. Sedov
 (Philadelphia:{\linebreak}
 {\hspace*{-0.6cm}}Soc.\ Ind.\ Appl.\ Math.), 568
\\[-0.57cm]
\item[\hspace{-0.3cm}]
Rigaut, F., Salmon, D., Arsenault, R., et al.\ 1998, PASP, 110, 152
\\[-0.57cm]
\item[\hspace{-0.3cm}]
Samarasinha, N. H. 2000, ApJ, 529, L107
\\[-0.57cm]
\item[\hspace{-0.3cm}]
Schr\"{a}pler,\,R., Blum,\,J., Seizinger,\,A., \& Kley,\,W. 2012, ApJ,~758,~35
% {\hspace*{-0.6cm}}35
\\[-0.57cm]
%
% \item[\hspace{-0.3cm}]
% Sekanina, Z. 1982, in Comets, ed. L. L. Wilkening (Tucson, AZ:
% {\hspace*{-0.6cm}}University of Arizona), 251
% \\[-0.57cm]
%
\item[\hspace{-0.3cm}]
Sekanina, Z. 1987, in Diversity and Similarity of Comets, ESA{\linebreak}
 {\hspace*{-0.6cm}}SP-278, ed. E. J. Rolfe \& B. Battrick (Noordwijk,
 Netherlands:{\linebreak}
 {\hspace*{-0.6cm}}ESTEC), 315
\\[-0.57cm]
%
% \item[\hspace{-0.3cm}]
% Sekanina, Z. 1988a, AJ, 95, 911
% \\[-0.57cm]
%
% \item[\hspace{-0.3cm}]
% Sekanina, Z. 1988b, AJ, 96, 1455
% \\[-0.57cm]
%
\item[\hspace{-0.3cm}]
Sekanina, Z. 1995, A\&A, 304, 296
\\[-0.57cm]
%
% \item[\hspace{-0.3cm}]
% Sekanina, Z. 1996, A\&A, 314, 957
% \\[-0.57cm]
%
\item[\hspace{-0.3cm}]
Sekanina, Z. 1998, ApJ, 509, L133
\\[-0.57cm]
\item[\hspace{-0.3cm}]
Sekanina, Z. 1999a, Earth Moon Plan., 77, 147
\\[-0.57cm]
\item[\hspace{-0.3cm}]
Sekanina, Z. 1999b, Earth Moon Plan., 77, 155
\\[-0.57cm]
%
% \item[\hspace{-0.3cm}]
% Sekanina, Z., \& Kracht, R. 2014, eprint arXiv:1404.5968
% \\[-0.57cm]
%
% \item[\hspace{-0.3cm}]
% Sekanina, Z., \& Kracht, R. 2015, ApJ, 801, 135
% \\[-0.57cm]
%
% \item[\hspace{-0.3cm}]
% Sekanina, Z., \& Kracht, R. 2016, eprint arXiv:1607.3440
% \\[-0.57cm]
%
\item[\hspace{-0.3cm}]
Sekanina, Z., \& Kracht, R. 2017, eprint arXiv:1703.00928 (Paper 1)
\\[-0.57cm]
%
% \item[\hspace{-0.3cm}]
% Sekanina, Z., Chodas, P. W., \& Yeomans, D. K. 1998, Plan.\ Space
%  {\hspace*{-0.6cm}}Sci., 46, 21
% \\[-0.57cm]
%
\item[\hspace{-0.3cm}]
Sosa, A., \& Fern\'andez, J. A. 2011, MNRAS, 416, 767
\\[-0.57cm]
%
% \item[\hspace{-0.3cm}]
% Szab\'o, Gy. M., Kiss, L. L., \& S\'arneczky, K. 2008, ApJ, 677, 121
% \\[-0.57cm]
%
\item[\hspace{-0.3cm}]
Szab\'o, Gy. M., S\'arneczky, K., \& Kiss, L. L. 2011, A\&A, 531, A11
\\[-0.57cm]
\item[\hspace{-0.3cm}]
Szab\'o, Gy. M., Kiss, L. L., P\'al, A., et al. 2012, ApJ, 761, 8
\\[-0.57cm]
\item[\hspace{-0.3cm}]
Vasundhara, R., \& Chakraborty, P.\ 1999, Icarus, 140, 221
\\[-0.57cm]
%
% \item[\hspace{-0.3cm}]
% Vsekhsvyatsky, S. K. 1958, Fizicheskie kharakteristiki komet.{\linebreak}
%  {\hspace*{-0.6cm}}(Moscow:\ Gosud.\ izd-vo fiz.-mat.\ lit.); translated: 1964,
%  Physical{\linebreak}
%  {\hspace*{-0.6cm}}Characteristics of Comets, NASA TT-F-80 (Jerusalem:\
%  Israel{\linebreak}
%  {\hspace*{-0.6cm}}Program for Scientific Translation)
% \\[-0.57cm]
%
\item[\hspace{-0.3cm}]
Weaver, H. A., \& Lamy, P. L. 1999, Earth Moon Plan., 79, 17
\\[-0.57cm]
\item[\hspace{-0.3cm}]
Weaver, H. A., Feldman, P. D., A'Hearn, M. F., et al.\ 1999, Icarus,{\linebreak}
 {\hspace*{-0.6cm}}141, 1
\\[-0.57cm]
\item[\hspace{-0.3cm}]
Whipple, F. L. 1950, ApJ, 111, 375
\\[-0.638cm]
\item[\hspace{-0.3cm}]
Williams, D. R., \& Wetherill, G. W. 1994, Icarus, 107, 117}
\\
%
% \item[\hspace{-0.3cm}]
% Yeomans, D. K., \& Chodas, P. W. 1989, AJ, 98, 1083
% \\
%
\vspace*{0.35cm} % 0.15cm
\end{description}
\end{document}